%% file: main.tex
\RequirePackage[bookmarksnumbered,unicode]{hyperref}

\documentclass[manuscript,screen]{acmart}

\usepackage{booktabs}
\usepackage{multirow}
\usepackage{array}
\usepackage{tikz}
\usepackage{soul}

\usepackage{amsmath,amssymb,amsfonts}
\usepackage{algorithmic}
\usepackage{graphicx}
\usepackage{textcomp}
\usepackage{xcolor}
 \usepackage{lstautogobble}
\usepackage{float}
\usepackage{listings}
 \usepackage{caption}
 \usepackage{subcaption}
 \usepackage{enumitem}
 \usepackage{moresize}

\definecolor{bluekeywords}{rgb}{0.13, 0.13, 1}
\definecolor{greencomments}{rgb}{0, 0.5, 0}
\definecolor{redstrings}{rgb}{0.75, 0, 0}
\definecolor{graynumbers}{rgb}{0.3, 0.5, 0.5}

\definecolor{background_ok}{rgb}{0.9, 0.9, 0.95}

\newcommand*\circled[1]{\tikz[baseline=(char.base)]{
            \node[shape=circle,draw,inner sep=1pt] (char) {#1};}}

\setcopyright{acmcopyright}
\copyrightyear{2026}
\acmYear{2026}
\acmDOI{XXXXXXX.XXXXXXX}

\acmJournal{CSUR}
\acmVolume{1}
\acmNumber{1}
\acmArticle{1}
\acmMonth{4}

\newcommand{\didem}[1]{{\color{blue}{\textbf{Didem}}: \color{red}{#1}}}
\newcommand{\flavio}[1]{{\color{orange}{\textbf{Flavio}}: {\color{orange}{#1}}}}

\newcommand{\ismayil}[1] {\color{orange}{\textbf{Ismayil}}: {\color{orange}{#1}}}

\newcommand{\edit}[1]{{\color{black}{#1}}}

\begin{document}

\title{The Landscape of GPU-Centric Communication}

\author{Didem Unat}
\affiliation{
\institution{Ko\c{c} University}
  \city{Istanbul}
  \country{Turkey}                    
}
\email{dunat@ku.edu.tr}    

\author{Ilyas Turimbetov}
\affiliation{
\institution{Ko\c{c} University}
  \city{Istanbul}
  \country{Turkey}                    
}
\email{iturimbetov18@ku.edu.tr}          
\author{Mohammed Kefah Taha Issa}
\affiliation{
\institution{Ko\c{c} University}
  \city{Istanbul}
  \country{Turkey}                    
}
\email{MISSA18@ku.edu.tr}    
\author{Doğan Sağbili}
\affiliation{
\institution{Ko\c{c} University}
  \city{Istanbul}
  \country{Turkey}                    
}
\email{dsagbili17@ku.edu.tr}    
\author{Flavio Vella}
\affiliation{
\institution{University of Trento}            
 \city{Trento}
  \country{Italy}                    
}
\email{flavio.vella@unitn.it}    
\author{Daniele De Sensi}
\affiliation{
\institution{Sapienza University of Rome}            
\city{Rome}
  \country{Italy}                    
}
\email{desensi@di.uniroma1.it}    
\author{Ismayil Ismayilov}
\affiliation{
\institution{Ko\c{c} University}
  \city{Istanbul}
  \country{Turkey}                    
}
\email{iismayilov21@ku.edu.tr}          
\renewcommand{\shortauthors}{D. Unat et al.}

\begin{abstract}
\input{sections/abstract}
\end{abstract}




\maketitle

\input{sections/introduction}

\input{sections/gpu-centric-intro}

\input{sections/vendor-mechanisms}

\input{sections/gpu-centric-paradigms}

\input{sections/outlook}

\input{sections/conclusion}

\section*{Acknowledgment}
Authors at Koç University were  supported from the European
Research Council (ERC) under the European Union’s Horizon 2020 research and innovation programme (grant agreement No 949587). Flavio Vella and Daniele De Sensi are supported by the European Union’s Horizon Europe under grant 101175702 (NET4EXA). Daniele De Sensi is supported by Sapienza University Grants ADAGIO and D2QNeT (Bando per la ricerca di Ateneo 2023 and 2024). 

\bibliographystyle{ACM-Reference-Format}
\bibliography{main}

\end{document}

%% file: sections/abstract.tex
In recent years, GPUs have become the preferred accelerators for HPC and ML applications due to their parallelism and high memory bandwidth. While GPUs boost computation, inter-GPU communication can create scalability bottlenecks, especially as the number of GPUs per node and cluster grows. Traditionally, the CPU managed multi-GPU communication, but advancements in GPU-centric communication now challenge this CPU dominance by reducing its involvement, granting GPUs more autonomy in communication tasks, and addressing mismatches in multi-GPU communication and computation.

This paper provides a landscape of GPU-centric communication, focusing on vendor mechanisms and user-level library supports. It aims to clarify the complexities and diverse options in this field, define the terminology, and categorize existing approaches within and across nodes. The paper discusses vendor-provided mechanisms for communication and memory management in multi-GPU execution and reviews major communication libraries, their benefits, challenges, and performance insights. Then, it explores key research paradigms, future outlooks, and open research questions. By extensively describing GPU-centric communication techniques across the software and hardware stacks, we provide researchers, programmers, engineers, and library designers insights on how to exploit multi-GPU systems at their best.

%% file: sections/introduction.tex
\section{Introduction}
In recent years, GPUs have become the accelerator of choice for a vast array of applications across both HPC and ML. This quick adoption, driven mainly by the GPU's massive parallelism and high memory bandwidth, means that most of modern cloud and HPC capability is now concentrated in clusters of GPUs. \edit{As of November 2025, 9 of the 10 leading Top500 supercomputers} rely on GPU clusters for acceleration and this trend is likely to continue \cite{top500}. 
The only system, Fugaku, without GPUs employs a highly vectorized CPU architecture combined with High-Bandwidth Memory.

While using scores of GPUs has been shown to significantly accelerate computation, communication between the GPUs can quickly become a scalability bottleneck \cite{Peta-scale-stencil} \cite{padal2}. Traditionally, multi-GPU communication, both within and across nodes, had always been the responsibility of the CPU. From their conception,  GPUs were thought of as devices that can supply large amounts of computation but that are also inherently dependent on the CPU for auxiliary tasks like communication. In this \textit{CPU-centric} model of execution, the routines relaying data for consumption by the GPUs are oblivious to the GPUs' existence.

In the last decade, several advancements, broadly referred to as \textit{GPU-centric} communication, have sought to challenge the CPU's hegemony on multi-GPU execution. At a high level, these advancements reduce the CPU's involvement in the critical path of execution, give the GPU more autonomy in initiating and synchronizing communication and attempt to address the semantic mismatch between multi-GPU communication and computation. 

In this paper, we present a comprehensive review of \textit{GPU-centric} communication with respect to vendor mechanism and user-level library supports. Our goal with this survey is to help 
allay the general state of confusion that arises when a prospective researcher begins wading into the field. We hope to help programmers, engineers, programming model and library designers 
understand 
the complexity and diversity of available options  because \textit{GPU-centric} communication spans a very broad spectrum of approaches, including hardware innovations like proprietary GPU-to-GPU interconnects and software mechanisms. These mechanisms have distinct benefits and challenges making it unclear when and where they should be preferred. The picture is made even murkier by inconsistent terminology used across literature and vendor differences in offerings.


We organize the paper as follows:
\begin{itemize}
    \item In Section \ref{Section:GPU-Centric-Intro}, we define the terminology, provide a definition for \textit{GPU-centric} communication. We taxonomize and abstract the existing approaches within a node and across nodes to eliminate any confusion.
    \item In Section \ref{Section:Vendor-Mechanisms}, we present the history and discuss  the vendor-provided mechanisms for enabling communication and networking, managing memory across devices in a multi-GPU execution. These mechanisms are used as building blocks for higher-level GPU-centric software libraries. 
    \item In Section \ref{Section:GPU-Centric-Libraries}, we list and compare the main communication libraries for both \textit{intra-} and \textit{inter-node} setups, discuss their benefits and challenges and rely on existing benchmarking works to provide insights into their performance.
    \item In Section \ref{Section:Outlook-and-Discussion}, we 
    discuss the main research paradigms underlying \textit{GPU-centric} communication,  provide an outlook on the field and present open research questions. 
\end{itemize}

\edit{Some methods and technologies discussed in this survey are inherently tied to proprietary ecosystems. While a substantial portion of the available literature and deployed systems focuses on NVIDIA technologies, we  present a balanced perspective by incorporating corresponding mechanisms from other vendors, such as AMD and Intel, whenever information is available. At the same time, several capabilities remain exclusive to specific vendors; we explicitly identify these vendor-specific features to ensure clarity and completeness.}




%% file: sections/gpu-centric-intro.tex
\section{Terminology and Communication Types} 
\label{Section:GPU-Centric-Intro}

We can loosely define \textit{GPU-centric communication} as \textit{mechanisms that reduce the involvement of the CPU in the critical path of multi-GPU execution}. This is a very broad definition covering a wide spectrum of solutions, involving both the vendor-level improvements that grant GPUs autonomy in communication and user-level implementations that leverage those improvements. To make this distinction clear, we discuss them in separate sections. In Section 3, we focus on communication mechanisms and primitives provided natively as part of the NVIDIA CUDA and AMD ROCm runtimes. \edit{In Section 4, we discuss how these mechanisms give rise to higher-level, GPU-centric communication libraries and present both vendor-provided software support by AMD, Intel and NVIDIA as well as other industry and academic solutions.}

We also point out the distinction between \textit{communication within a node (intra-node)} and \textit{communication across nodes (inter-node)}. A single GPU-accelerated node comprises a single shared memory host with multiple GPU cards attached. When communicating within a node, any given GPU can be controlled by a single thread or process, with a shared memory and address space. A multi-node system has multiple such nodes where a different process controls each GPU, and memory is not shared between processes running on different nodes. The communication landscape changes depending on the setup used, since inter-node communication requires handling the GPU-NIC interaction and across-process communication.


\subsection{Intra-Node Communication}
Despite classification of communication methods into GPU- and CPU-side is commonly used and can be sufficient for an end-user, it is not always explanatory and accurate. 
To avoid the vagueness of definitions, we divide the communication methods into several types. 
These types are based on {\em the executor of each of the operations performed during communication}. 
We define two main operations needed for the communication to take place in the intra-node scenario and four operations for the inter-node one.
In the intra-node case, the two components of a communication call are:
\begin{itemize}
    \item \textbf{API}. Defines where the communication API call is made by the programmer or library.  
    \item \textbf{Data path}. Indicates who participates in the data movement and shows the corresponding data path.  
\end{itemize}
The examples showing the classification of intra-node communication mechanism, together with figures depicting the data paths are shown on Table \ref{table:intra_types} and Figure \ref{fig:intra_types}.

\input{listings/intra-node-table-fig}

The communication method in \circled{1}, referred as {\em host native}, is made on the host side and does not involve direct P2P (peer-to-peer) access between the devices.  
These involve all methods that can be launched on the host side with P2P access disabled. 
Otherwise, as \circled{2} {\em (host-controlled)} shows, the communication does not involve an extra copy to the host memory, and passes directly through PCIe, NVLink or Infinity Fabric interconnects. 
GPUCCL (i.e., NCCL, RCCL, oneCCL, etc), GPU-aware MPI and *memcpy operations have host-side API, so they may belong to both \circled{1} and \circled{2}. 
By enabling direct access to peer device memory, the device-side API eliminates CPU involvement from both the data and control paths in intra-node communication, as illustrated by \circled{3}, referred to as device native, in Figure~\ref{fig:intra_types}. 
NVSHMEM, ROCSHMEM, \edit{Intel SHMEM} offer host-side API as well, but require P2P access, so their host-side APIs belong to \circled{2} and device-side API is of type \circled{3}. In-kernel P2P direct load and stores offer similar functionality and belong to type \circled{3}, but work even with P2P access disabled, which is type \circled{4}, where data path falls back to the host.

\subsection{Inter-Node Communication}

The inter-node scenarios are more diverse since interaction with the NIC has to take place. The details of each method's implementations may involve complicated data paths and decision making. We distinguish four main components of inter-node communication for a simpler classification. Apart from the API and the data path (to the NIC) used in the intra-node scenario, there are two additional components involving interaction with the NIC:

\begin{itemize}
    \item \textbf{Register}/construct messages. This step involves the construction of data packets and their registration on the NIC.
    \item \textbf{Trigger} communication. It defines who rings the doorbell on the NIC to issue data transfer.
\end{itemize}

\input{listings/inter-node-table-fig}

We identify five main categories when classifying inter-node communication. From \circled{1} to \circled{5}, as Table \ref{table:inter_types} and Figure \ref{fig:inter_types} show, more components of communication calls were moved to the GPU side, while the data transfer path saw a reduction in the number of data copies required to reach the NIC. Over the years, the communication methods and the corresponding technologies  depict the optimizations of both data transfer and communication control, which will be discussed in Section \ref{Section:Vendor-Mechanisms}. First, entirely CPU-side communication method in \circled{1} was available, which has been improved by removal of extra copy between CPU-GPU and CPU-NIC buffers in \circled{2} \edit{with the help of shared pinned memory between GPU and NIC. This reduced the latency for GPU-NIC transfers.} \edit{After that, GPU RDMA (Remote Direct Memory Access) \circled{3} facilitated direct access to the GPU memory from NIC over PCIe minimizing the data path between them. }\circled{4} represents GPU-triggered communication technologies such as GPUDirect Async and GPU-TN \cite{GPU-TN-GPU-Triggered-Networking-for-Intra-Kernel-Communication}, where the GPU has become capable to initiate communications, given that CPU prepared the packets on the NIC in advance. \circled{5} moves the packet preparation and interaction with the NIC entirely to the GPU as well, making device-native communication possible. 

The types given in Table \ref{table:inter_types} and Figure \ref{fig:inter_types} do not reflect all possible combinations, since some libraries, based on the configuration and the available hardware may lead to different combinations of data paths and control. For example, without the RDMA technology, even with GPU-side communication control, the data path will involve the host memory.
\edit{For instance, Intel SHMEM uses a proxy thread on the host to execute actual RDMA using a standard OpenSHMEM library even though the communication call is issued within a device kernel running on the GPU. }

%% file: listings/intra-node-table-fig.tex
\begin{figure}
    \begin{tabular}{|m{8em}|m{3em}|m{4em}|m{22em}|}
    \hline
    Type & API & Data Path & Examples \\
    \hline
    \circled{1} Host Native & Host & Host & cuda/hipMemcpy and other host-side comm. (No P2P) \\
    \hline
    \circled{2} Host-Controlled & Host & Device & cuda/hipMemcpy and other host-side comm. (P2P) \\
    \hline
    \circled{3} Device Native & Device & Device & NVSHMEM / ROCSHMEM / \edit{Intel SHMEM,} GPU-side, direct load/store (P2P) \\
    \hline
    \circled{4} Host Fallback & Device & Host & Direct load/store (No P2P) \\
    \hline
    \end{tabular}
    \captionof{table}{Types of intra-node communication methods}
    \label{table:intra_types}
\end{figure}
\begin{figure}
    \includegraphics[width=\linewidth]{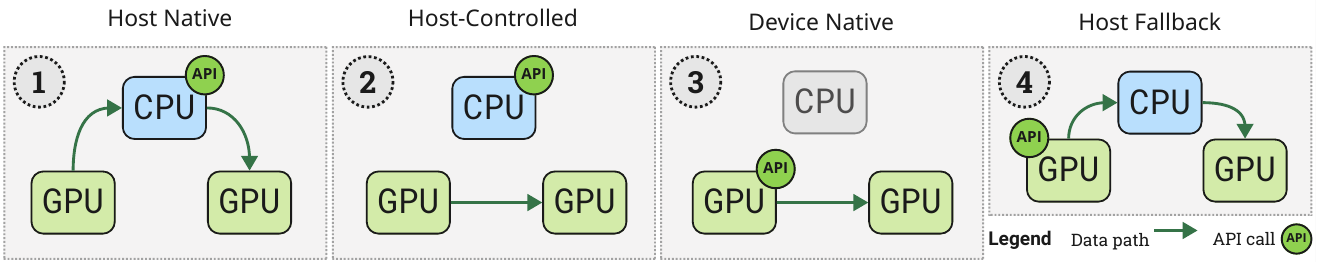}
    \captionof{figure}{Data paths and API calls of intra-node communication methods. Figure is available under CC-BY \cite{figure-cc-by}.}
    \label{fig:intra_types}
\end{figure}

%% file: listings/inter-node-table-fig.tex
\begin{figure}
    \begin{tabular}{|m{9em}|m{3em}|m{3em}|m{3em}|m{10em}|m{10em}|}
    \hline
    Type & API & Register & Trigger & Data path & Examples \\
    \hline
    \circled{1} \edit{Host Native}  & Host & Host & Host & Through host (2 copies) & GPU-Aware MPI before \circled{2} \\
    \hline
    \circled{2} \edit{Pinned Host Native} & Host & Host & Host & \textbf{Through host (1 copy)} & GPUDirect 1.0 in GPU-Aware MPI, NCCL, \edit{RCCL, oneCCL} \\
    \hline
    \circled{3} \edit{GPU RDMA}& D/H & Host & Host & \textbf{Direct} & GPUDirect RDMA, ROCmRDMA in GPU-Aware MPI, NCCL, NVSHMEM, \edit{ROCSHMEM} \\
    \hline
    \circled{4} \edit{GPU-Triggered}& D/H & Host & \textbf{Device} & Depends on \circled{3} & GPUDirect Async in GPU-Aware MPI, \edit{NCCL v2.28}, NVSHMEM, \edit{ROCSHMEM} \\
    \hline
    \circled{5} Device Native & Device & \textbf{Device} & Device & Depends on \circled{3} & NVSHMEM with IBGDA, GPUrdma \\
    \hline
    \end{tabular}
    \captionof{table}{Types of inter-node communication methods. Cells in bold refer to where a change or optimization has been made. \textit{D/H} means that both device-side and host-side API calls may belong to this type. }
    \label{table:inter_types}
\end{figure}

\begin{figure}
\centering
    \includegraphics[width=\linewidth]{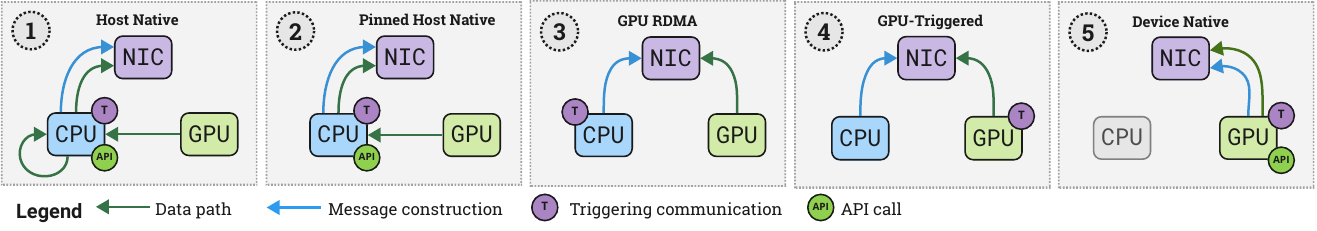}
    \captionof{figure}{Inter-node communication data and control paths. Figure is available under CC-BY \cite{figure-cc-by}.}
    \label{fig:inter_types}

\end{figure}

%% file: sections/vendor-mechanisms.tex
\section{Vendor Mechanisms} \label{Section:Vendor-Mechanisms}
\begin{figure*}[th]
      \includegraphics[width=\linewidth]{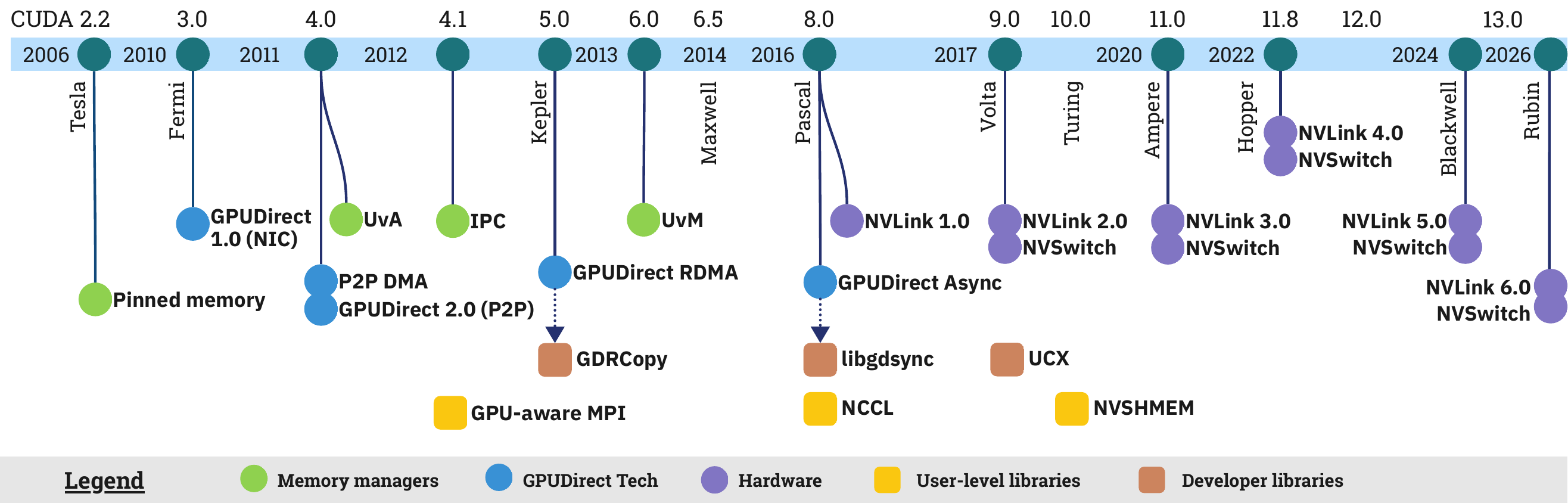}
    \caption{Timeline of NVIDIA technologies enabling GPU-centric communication and networking. Figure is available under CC-BY \cite{figure-cc-by}.}
    \label{fig:timeline}
\end{figure*}

In this section we discuss the vendor-provided mechanisms for enabling communication and  networking, managing memory across devices in a multi-GPU execution. These mechanisms are provided by GPU programming model runtimes or as part of the extended APIs. The technologies are classified into four categories: memory managers, GPUDirect technologies, hardware, and libraries. 
Figure \ref{fig:timeline} summarizes the technologies provided by NVIDIA, detailing their timeline and availability. 

Next, we introduce the memory management mechanisms and GPUDirect technologies, followed by the hardware support that served as precursors and ultimately made these communication methods viable. These technologies form the backbone of the higher-level GPU-centric libraries, which are discussed in Section 4.

\subsection{Memory Management Mechanisms}

\subsubsection{\textbf{Page-Locked / Pinned Memory}}
By default, memory allocated on the host using device malloc (\textit{e.g. cudaMalloc(), hipMalloc()} is \textit{pageable} and is not accessible to the GPU. When a transfer between pageable host memory and device memory is performed, the GPU runtime must first stage the host data through a temporary buffer in \textit{page-locked memory} and then copy the data from page-locked memory to the GPU. To avoid the pageable $\rightarrow$ page-locked memory copy, \textit{cudaMallocHost()} allows allocating page-locked memory directly, skipping the intermediate copy stage. 
Because of this, page-locked memory is also referred to as \textit{zero-copy} or \textit{pinned} memory \cite{CUDA-Programming-Guide-Release-12.2, CUDA-Blog-Pinned-Memory-How-to-Optimize-Data-Transfers-in-CUDA-C/C++}. 

Pinned memory, known for its high bandwidth and low latencies in host-device transfers \cite{Evaluating-Characteristics-of-CUDA-Communication-Primitives-on-High-Bandwidth-Interconnects}, efficiently coordinates CPU-GPU execution by enabling direct system-wide access. It is also utilized with GPUDirect RDMA for improved inter-node communication \cite{Evaluating-Modern-GPU-Interconnects}. However, its physical memory locking can lead to high memory consumption, potentially impacting system performance with excessive allocations \cite{CUDA-Blog-Pinned-Memory-How-to-Optimize-Data-Transfers-in-CUDA-C/C++}.



\subsubsection{\textbf{Unified Virtual Addressing (UVA)}}
UVA is a memory management technique  which allows all GPUs and CPUs within a node to share the same unified virtual address space \cite{CUDA-Programming-Guide-Release-12.2, CUDA-4.0-Release-Notes}. Prior to UVA, host $\leftrightarrow$ device and device $\leftrightarrow$ device copies had to explicitly specify the direction of transfer. With UVA, the physical memory location can be inferred from pointer values, thus, reducing the overhead of managing separate memory spaces and enabling libraries to simplify their interfaces \cite{Talk-Peer-to-Peer-and-Unified-Virtual-Addressing}.

\subsubsection{\textbf{IPC}} 
In early GPU runtimes versions, pointers could not be accessed across process boundaries, so memory copies between GPU buffers had to go through the host, creating a bottleneck. To overcome this limitation, Inter-Process Communication (IPC) enables processes on the same node to access device buffers of other processes without additional copies \cite{CUDA-4.1-Release-Notes}. With  IPC, memory handles are created and passed between processes using standard IPC mechanisms, resulting in lower latencies than staging copies through the host. 



\subsubsection{\textbf{Unified Virtual Memory (UVM)}}
UVM allows for the allocation of \textit{managed} memory through \textit{cudaMallocManaged()} calls by creating a single address space accessible to all processors within a single node. UVM works by dividing the requested memory into pages that are resident on the CPU. The programmer can access memory on a device without explicit copies. If a memory access is part of a page that is not on the device, the UVM-driven triggers a page-fault that automatically migrates the page to the requesting device. The UVM driver can also evict pages from a given device back to host memory when the total paged memory size exceeds device memory \cite{CUDA-Programming-Guide-Release-12.2}.

UVM provides several benefits in regard to programmability. First, programmers are exposed to a single unified address space that they can access as if the whole allocated chunk of memory is resident on a single GPU. Any copies occurring around the system are implicit and hidden from the programmer's view. Additionally, UVM allows memory oversubscription whereby more memory can be allocated than all the multi-GPU device memory combined. This is possible since most of the memory can stay on the CPU and be paged in whenever a given device requests it \cite{CUDA-Blog-Improving-GPU-Memory-Oversubscription-Performance, Oversubscribing-GPU-Unified-Virtual-Memory-Implications-and-Suggestions}.





\subsection{GPUDirect Technologies}


\subsubsection{\textbf{GPUDirect 1.0 (NIC)}}
 GPUDirect 1.0 allowed GPUs and NICs to share the same pinned memory region. Prior, the pinned memory regions in system memory for GPUs and the NIC were separate. By implication, to communicate GPU data across nodes, the GPU first copied the data to its pinned memory region, the CPU then copied it to the NIC's memory region, only then it can be accessed by the NIC, which sent it across the network, as shown in Figure \ref{fig:inter_types} \circled{1}. The intermediate CPU-initiated copy from GPU $\rightarrow$ NIC pinned memory regions adds CPU overhead and increases the latency for GPU communication. GPUDirect 1.0 introduced a shared memory GPU-NIC pinned memory region, thus, avoiding the intermediate CPU-initiated copy \cite{Talk-State-Of-GPUDirect-Technologies, GPUDirect-1.0-The-development-of-Mellanox/NVIDIA-GPUDirect-over-Infiniband}. 

\subsubsection{\textbf{GPUDirect 2.0 (Peer-to-Peer)}}
Along with the introduction of UVA, the CUDA 4.0 release added support for direct peer-to-peer communication among GPUs in a single node that share the same PCIe root complex \cite{CUDA-4.0-Release-Notes}. This functionality was encapsulated in a technology known as GPUDirect 2.0 or GPUDirect P2P. Instead of staging data through the host, GPUs could now directly access each other's memory over PCIe, establishing, for the first time, a direct GPU-to-GPU data path. These changes led to two new communication mechanisms: \textit{P2P DMA Copies} whereby a \textit{cudaMemcpy} call would trigger a DMA transfer directly between source and target GPU memories and \textit{P2P Direct Load / Stores} using which the GPUs could directly access data by dereferencing pointers to the remote GPU buffers. GPUDirect P2P also added support for NVLink (Section \ref{sec:NVLink}) when the latter technology was introduced \cite{NVIDIA-GPUDirect, Talk-NVIDIA-GPUDirect, Talk-State-Of-GPUDirect-Technologies}.

GPUDirect P2P provided two main benefits. It eliminated redundant GPU $\leftrightarrow$ CPU copies and host buffers, which were required when the transfers were staged through the CPU. Also, by eliding the need to maintain communication buffers on the host and providing a new communication mechanism (P2P Direct Load / Stores), GPUDirect P2P increased the convenience of multi-GPU programming \cite{Talk-NVIDIA-GPUDirect}.

We note that P2P DMA Copies can also work without UVA support. If UVA is not enabled, P2P DMA Copies can be performed using the \textit{cudaMemcpyPeer()} variants by explicitly specifying the target GPU. However, P2P Direct Load / Stores will not work without UVA as directly accessing a remote GPU's pointer presumes a unified address space \cite{CUDA-Programming-Guide-Release-12.2}.

\subsubsection{\textbf{GPUDirect RDMA}} \label{GPUDirect RDMA}
With the introduction of GPUDirect RDMA in CUDA 5.0, direct communication between NVIDIA GPUs across nodes became feasible. GPUDirect RDMA facilitates a direct communication channel between GPUs and third-party devices through standard PCIe features. The technology exposes segments of GPU memory on the PCIe memory resource, referred to as the Base Address Register (BAR) region. This enables NICs to directly read/write GPU memory without routing through the host \cite{GPUDirect-RDMA}. Analogously, AMD offers ROCm RDMA (previously called ROCnRDMA) \cite{ROCnRDMA, ROCK-Kernel-Driver}.
GPUDirect RDMA provides several optimizations to the data path, namely by eliminating additional copies to host memory, reducing inherent latencies stemming from GPU-NIC interaction, increasing bandwidth and reducing CPU overhead. 

\subsubsection{\textbf{GPUDirect Async}} While previous GPUDirect technologies focused on improving the data path, GPUDirect Async optimizes the control path between the GPU and the NIC. Introduced in CUDA 8.0, it enables GPUs to initiate and synchronize network transfers, thereby reducing the CPU's involvement in the critical path. GPUDirect Async works by having the CPU pre-register messages, which the GPU kernel can then trigger by ringing a doorbell on the NIC. As a result, the GPU can continue executing while the communication is being triggered, rather than needing to stop for the CPU to initiate the communication, as was previously necessary \cite{GPUDirect-Async-1-Offloading-Communication-Control-Logic-in-GPU-Accelerated-Applications, GPUDirect-Async-2-Exploring-GPU-Synchronous-communication-techniques-for-Infiniband-clusters}.

Although GPUDirect Async has led to improvements in efforts to move the control path away from the CPU, it still does not completely transfer the control path to the GPU since communication is limited to kernel launch boundaries. Essentially, the GPU can only initiate messages previously registered by the CPU. Further improvements to GPUDirect Async are implemented as part of the IBGDA transport in the NVSHMEM library (Section \ref{NVSHMEM}).

\subsection{GPUNetIO}
GPUNetIO~\cite{Blog-Inline-GPU-Packet-Processing-NVIDIA-DOCA-GPUNetIO} is a technology solution proposed by NVIDIA as part of DOCA (Datacenter-On-a-Chip Architecture)~\cite{NVIDIA-DOCA-SDK}. DOCA is a full-stack software framework designed to facilitate the development of applications for NVIDIA BlueField Data Processing Units (DPUs)~\cite{Nvidia-Data-Center-DPU-Architecture}. On non-RDMA networks GPUNetIO allows the GPU to send, receive, and process network packets~\cite{Blog-Optimizing-Inline-Packet-Processing-DPDK-GPUDev-GPUs, Blog-Inline-GPU-Packet-Processing-NVIDIA-DOCA-GPUNetIO}. On RDMA networks (both RoCE and InfiniBand)~\cite{NVIDIA-DOCA-SDK}, from DOCA v2.7, GPUNetIO allows the GPU to execute RDMA send and receive \emph{not only on kernel boundaries}, but at any point during the kernel execution. In a nutshell, GPUNetIO allows the GPU to interact with the NIC without any CPU intervention.

On RDMA networks, the GPU kernel can wait (in blocking or nonblocking mode) for the completion of RDMA receive operations. On non-RDMA networks, GPUNetIO provides semaphores that can be used explicitly within the kernel for synchronization with the NIC when sending and receiving packets. Semaphores can also be used to synchronize GPU kernels with the CPU (in case the packet processing is split between the CPU and the GPU) or with other CUDA kernels (if the packet processing is split across multiple kernels).

\subsection{\textbf{Modern GPU-centric Interconnects}} \label{sec:NVLink}
\edit{GPU-centric interconnect technologies provide high-bandwidth, low-latency communication between multiple GPUs within a node, a capability critical for high-performance workloads, particularly in AI training and HPC. }

NVLink is a proprietary interconnect technology for NVIDIA GPUs. Its design addresses the bandwidth limitations of PCIe, which has been observed to be a transfer bottleneck in GPU-accelerated applications \cite{Evaluating-Modern-GPU-Interconnects, DGX-1-White-Paper}. Table \ref{table:NVLink_Generation_Spec} presents the specifications of each generation of NVLink \cite{Pascal-GPU-NVLink, NVLink-and-NVSwitch, Evaluating-Modern-GPU-Interconnects}.

\begin{table} [htbp]
    \centering
    \begin{tabular}{|m{5em}|m{5em}|m{7em}|m{8em}|m{6em}|}
\hline
Generation & Number of NVLink slot & Per direction Bandwidth per NVLink (GB/sec) & Total Aggregate Bi-Directional Bandwidth (GB/sec) & Supported \newline Architecture\\
\hline
First & 4 & 20 & 160 & Pascal \\
\hline
Second & 6 & 25 & 300 & Volta \\
\hline
Third  & 12 & 25 & 600 & Ampere \\
\hline
Fourth & 18 & 25 & 900 & Hopper \\
\hline
Fifth & 18 & 50 & 1800 & Blackwell \\
\hline
\end{tabular}
    \caption{NVLink Generation Specifications}
    \label{table:NVLink_Generation_Spec}
\end{table}

Additionally, NVLink was also used to connect GPUs with the CPU for IBM Power8 and Power9 CPUs but with the introduction to Grace Hopper Superchip, NVLink is used as a Chip-to-Chip (C2C) interconnect with 900 GB/sec bi-directional bandwidth \cite{Evaluating-Modern-GPU-Interconnects,NVIDIA-Grace-Hopper}. Later with the introduction of Grace Blackwell Superchip, NVLink-C2C is used for connecting Grace CPU with 2 Blackwell GPUs with a total of 3.6 TB/sec bidirectional bandwidth\cite{NVIDIA-GB200}. 
Fifth-generation NVLink on NVIDIA Blackwell delivers  1.8TB/s bidirectional throughput per GPU, providing high-speed communication among up to 576 GPUs.

The introduction of NVLink optimized the bandwidth between NVIDIA GPUs, turning P2P communication into a viable mechanism for intra-node communication and shifting the \textit{data path} heavily in favor of GPUs. A disadvantage of NVLink is that it is not self-routed meaning that if any two given GPUs do not have a direct NVLink connection communication will have to be routed through an intermediate GPU \cite{Evaluating-Modern-GPU-Interconnects}. This limitation is overcome by NVSwitch \cite{NVLink-and-NVSwitch}, a backboard technology that can implement all-to-all connections between all GPUs. As an example, a DGX-2 node consists of 16 V100 GPUs that are all-to-all connected through NVLink and NVSwitch \cite{DGX-2}. Starting from the third generation, NVSwitch supports SHARP \cite{nvswitch-sharp}, which offloads allreduce operations to NVSwitch, allowing allreduce to operate at full line rate \cite{nvswitch-slides}.


\edit{AMD’s alternative is Infinity Fabric/xGMI, used in modern AMD GPU accelerators (e.g., the AMD Instinct MI300X). xGMI provides high-bandwidth 
communication between GPUs within a node. 
The architecture supports an aggregate bidirectional inter-GPU bandwidth of 896 GB/s for an 8-GPU system. Unlike NVSwitch, AMD’s current interconnect mesh is flat and does not rely on a switching fabric. This places some limits on scaling beyond 8 GPUs per node. Recently, the Ultra Accelerator Link (UALink) consortium was established to develop a more open shared memory accelerator interconnect, compatible with multiple technologies and vendors \cite{ualink}.

Intel’s Xe-Link fabric is the high-bandwidth, fully connected intra-node interconnect used in systems such as Aurora supercomputer, linking the Intel Data Center GPU Max (Ponte Vecchio) devices into a unified topology. In typical Aurora nodes, six GPUs are connected in an all-to-all configuration, though Xe-Link can also support up to 8-way arrangements where every GPU has direct links to all others. These links support load/store accesses, copy-engine transfers, and remote atomic operations, enabling GPUs to directly access each other’s memory without host. 
}

\input{sections/discussion_onvendor_mech}

%% file: sections/discussion_onvendor_mech.tex
\subsection{Discussion on Vendor Mechanisms}

\subsubsection{{\bf Impact of GPUDirect P2P and Direct Load/Store on Programming}}

The introduction of GPUDirect P2P marked a significant shift in the paradigm of multi-GPU execution, enabling direct communication between GPUs using load and store operations from within the kernel. 
Direct Load/Store-based communication offers several benefits. First, it allows the programmer to inline communication with computation.
The programmer no longer has to rely on separate models for communication and computation and can instead combine them within the GPU kernel \cite{NVSHMEM-Exploring-OpenSHMEM-Model-to-Program-GPU-Based-Extreme-Scale-Systems, NVSHMEM-Efficient-Breadth-First-Search-on-Multi-GPU-Systems-Using-GPU-Centric-OpenSHMEM, Talk-NVSHMEM-GPU-Integrated-Communication-for-NVIDIA-GPU-Clusters}. Second, Direct Load/Stores utilize the high levels of parallelism offered by the GPU and can achieve higher levels of bandwidth and lower latencies compared to DMA copies \cite{Groute, Interconnect-Bandwidth-Heterogeneity-on-AMD-MI250x-and-Infinity-Fabric}. Third, Direct Load/Stores can \textit{implicitly} overlap communication with computation through the GPU's inherent latency hiding capabilities. Given both the high levels of parallelism granted by the GPU and the increasing bandwidth numbers offered by modern interconnects, the GPU has the capability to hide latencies not only to local but remote memory as well \cite{NVSHMEM-Exploring-OpenSHMEM-Model-to-Program-GPU-Based-Extreme-Scale-Systems, NVSHMEM-Efficient-Breadth-First-Search-on-Multi-GPU-Systems-Using-GPU-Centric-OpenSHMEM, NVSHMEM-GPU-Centric-Communication-on-NVIDIA-GPU-Clusters-with-Infiniband-A-Case-Study-with-OpenSHMEM}. This is another boon for the programmer as the method of achieving overlap is shifted from a manual software-based approach implemented by the programmer through streams and events to an automatic hardware-based overlap. Since the onus of communication/computation overlap is passed from the programmer to the hardware, another implication is that the overlap will improve as the hardware gets better at hiding memory latencies. Fourth, Direct Load / Stores expand the scope of applications that could be accelerated through multiple GPUs. Traditionally, applications with fine-grained communication patterns achieved poor scalability on multi-GPU systems as computation frequently had to be interrupted and synchronized in order for the CPU to initiate communication. With Direct Load / Stores from within the kernel, GPUs can adapt well to fine-grained communication patterns. 
\edit{Finally, Direct Load / Stores allows communication to be triggered within the kernel without leaving the GPU. This direction is particularly promising when combined with persistent kernels \cite{Multi-GPU-Communication-Schemes-for-Iterative-Solvers-When-CPUs-Are-Not-in-Charge, PERKS-A-Locality-Optimized-Execution-Model-for-Iterative-Memory-Bound-GPU-Applications}, where a single kernel is launched and maintains its execution across multiple work iterations by employing an internal loop on the device, thereby minimizing kernel launch overhead.}
\edit{In fact, Spector et al. used direct Load/Store on tensor-parallel inference with Llama-70B and implemented a megakernel to perform asynchronous communication, overlapping them with compute and local memory operations \cite{weboughtthewholegpu}.}


Despite the improvements conferred by Direct Load / Stores, there are several inherent challenges. First, a fundamental challenge is that communication and computation contend for the same limited resource as they now both require large volumes of GPU threads to make progress. This can be especially problematic when communication is implemented as a separate kernel. If the computation kernel is launched first, it can potentially monopolize all GPU resources preventing the communication kernel from being launched, effectively, eliminating any possibility of overlap. It is possible to alleviate this issue by launching the communication stream with a higher priority so that it is always scheduled first. We note that P2P DMA Copies do not have this issue as they use the GPU's DMA / Copy Engines - a physically separate resource - for communication \cite{Benchmarking-multi-GPU-Applications-on-Modern-multi-GPU-integrated-systems}. Second, similar to single-GPU memory accesses, P2P Direct Load / Stores are highly sensitive to memory coalescing with random non-coalesced access performing far worse than coalesced accesses \cite{Groute}. Such non-coalesced Direct Reads may expose remote memory latencies which are beyond the GPU scheduler's ability to hide, eventually, stalling execution. On a similar note, sporadic non-coalesced Direct Writes at sub-cacheline granularities may dramatically underutilize the interconnect \cite{PROACT}.

\subsubsection{{\bf Limitation of GPUDirect RDMA}}
A significant limitation of GPUDirect RDMA is that there are no guarantees of consistency between GPU and NIC memories \textit{while a kernel is running}. Consistency is guaranteed only by returning control to the CPU by tearing down the kernel and launching a new kernel, thus, limiting communication to kernel boundaries. This also implies that combining persistent kernels with GPU-initiated inter-node communication will inevitably lead to data correctness issues \cite{GPUDirect-RDMA}. Chu et al. get around this limitation by issuing a PCIe read from the NIC to GPU memory which flushes the previous NIC writes to the GPU and guarantees memory ordering \cite{Designing-High-Performance-In-Memory-Key-Value-Operations-with-Persistent-GPU-Kernels-and-OpenSHMEM}. Since version 11.3, CUDA also offers the  \textit{cudaDeviceFlushGPUDirectRDMAWrites()} API which can be used to enforce consistency similarly \cite{CUDA-Runtime-Device-Management-Flush-GPUDirect-RDMA-Writes, Talk-GTC21-Latest-in-GPUDirect}. While useful, CUDA still relies on the CPU to enforce GPU-NIC consistency. AMD, on the other hand, has explicitly corrected the GPU-NIC consistency issues in the context of device-side communication from persistent kernels and integrated the proposed fixes into ROC\_SHMEM \cite{GIO-GPU-Initiated-OpenSHMEM-Correct-and-Efficient-Intra-Kernel-Networking-for-dGPUs}. We further discuss this issue in the context of CPU-free networking in Section \ref{CPU-Free Networking}.

\input{sections/Gpu-triggered}

%% file: sections/Gpu-triggered.tex
\subsubsection{\bf {Enabling Triggering Capability in GPU-Centric Communication}}


In type \circled{3} (GPUDirect/ROCn RDMA) introduced in Section \ref{Section:GPU-Centric-Intro}, the CPU is still responsible for the initial configuration of the system,  the data transfer preparation and initiating transfers. 
The first phase includes setting up network interfaces and loading GPU drivers. The CPU registers GPU memory with the RDMA-capable NIC. This registration by host allows the NIC to directly access GPU memory, bypassing the need for intermediary CPU steps during data transfer.
In phase two, the CPU allocates GPU memory buffers and ensures they are aligned correctly. These buffers will be used for efficient data transfer to and from the GPU.
Then, the CPU sets up GPU streams and events that manage and sequence the data transfers and, guarantees the compilation of the work. Streams are used to queue operations, ensuring they are executed in the correct order.
However, for truly low-latency applications, this cost may still represent a bottleneck as the mechanism 
relies on several synchronization points through the streams~\cite{Exploring-GPU-Stream-Aware-Message-Passing-using-Triggered-Operations, Exploring-Fully-Offloaded-GPU-Stream-Aware-Message-Passing, Optimizing-Distributed-ML-Communication-with-Fused-Computation-Collective-Operations}.  

GPU-trigger communication in type \circled{4}  and \circled{5} defined in Section \ref{Section:GPU-Centric-Intro} facilitates the offloading of both computation and communication control paths to the GPU by removing the synchronization cost described above.
Here, triggered operations play a crucial role, as they are special tasks scheduled to execute only when specific conditions are met. In the stream-triggered (ST) strategy, these operations manage data transfers and synchronization through the GPU control processor, thereby reducing CPU involvement.
Deferred Execution is another essential aspect, where the CPU creates command descriptors with deferred execution semantics and appends them to the NIC command queue. These descriptors are executed when the conditions specified by the GPU control operations are fulfilled. For example, in the HPE Slingshot 11 NIC supports these deferred operations~\cite{Exploring-Fully-Offloaded-GPU-Stream-Aware-Message-Passing}\footnote{2023. Libfabric Deferred Work Queue, \url{https://ofiwg.github.io/libfabric/v1.9.1/
man/fi_trigger.3.html}}, including sending and receiving communications that are triggered when a hardware counter reaches a given threshold. This is possible by enabling specific command queues (e.g.,  Libfabric Deferred Work Queues). 
The synchronization between the GPU control processor and the NIC ensures the successful completion of communication operations through a special mechanism (e.g., in NVIDIA DOCA this is implemented with GPUNetIO semaphores~\cite{Blog-Inline-GPU-Packet-Processing-NVIDIA-DOCA-GPUNetIO}). 

\begin{figure}[t]
    \centering
    \includegraphics[width=0.6\linewidth]{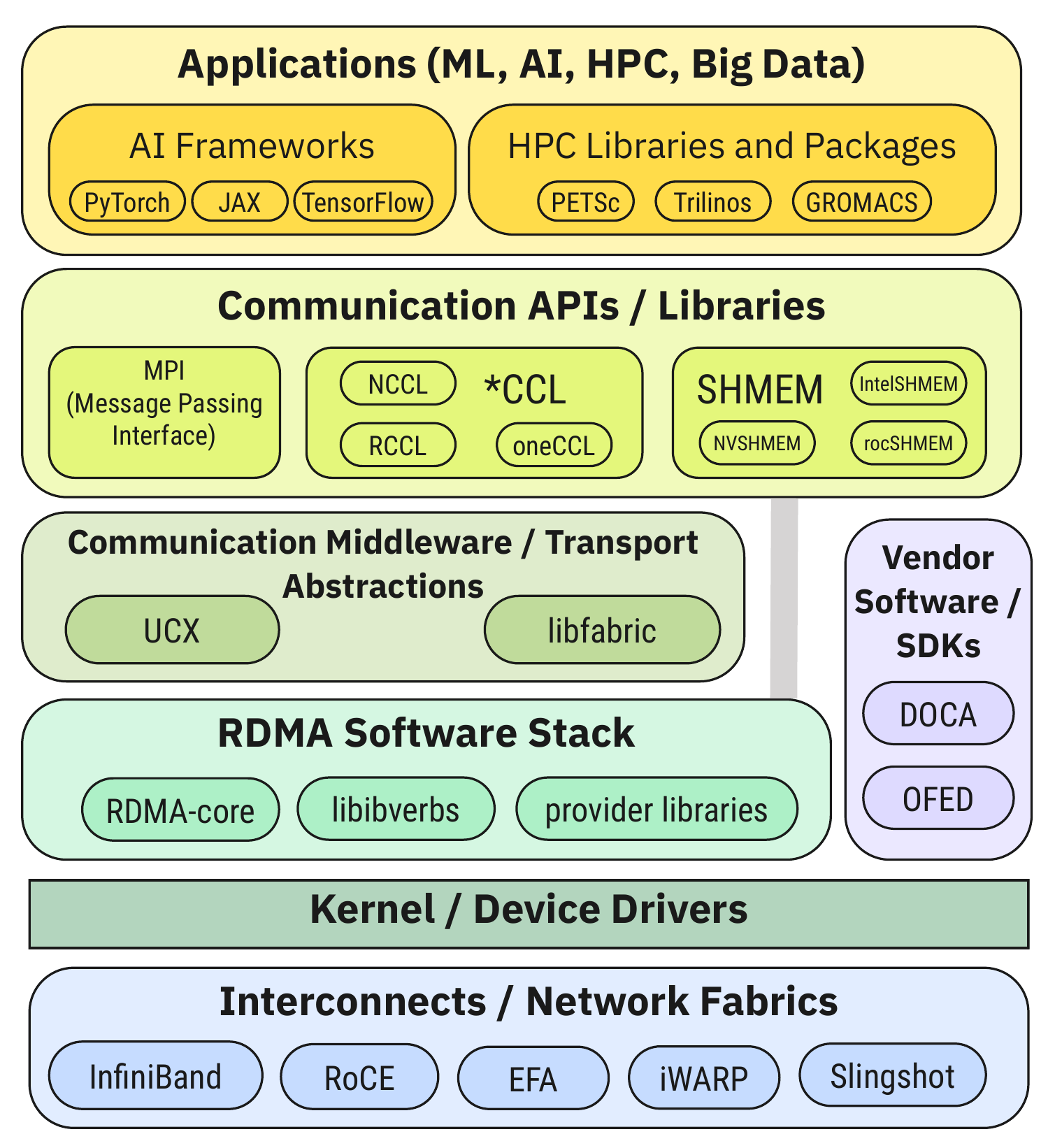}
    \caption{RDMA-oriented software stack for GPU-centric communication, showing the layers from applications and domain frameworks to communication libraries, transport middleware, provider libraries, kernel drivers, network fabrics, and vendor software stacks. Figure is available under CC-BY \cite{figure-cc-by}.}
    \label{fig:networkstack}
\end{figure}

To contextualize these vendor mechanisms within the broader communication ecosystem, Figure \ref{fig:networkstack} summarizes the software stack from applications and communication libraries down to middleware, drivers, and network fabrics. This layered view helps connect the low-level mechanisms discussed in this section with the user-level communication libraries reviewed next.

%% file: sections/gpu-centric-paradigms.tex
\section{GPU-Centric Communication Libraries} 
\label{Section:GPU-Centric-Libraries}
We now discuss the main GPU-centric communication libraries namely, GPU-aware MPI, GPU-centric Collectives and GPU-centric OpenSHMEM that have sprouted in recent years to ease multi-GPU programming.

\subsection{GPU-Aware MPI}
Given that MPI is the de facto \textit{lingua franca} of HPC, much effort has gone into making it interoperable with GPU programming models, culminating in \textit{GPU-Aware MPI} implementations that \textit{can differentiate between host and device buffers}.
Prior to \textit{GPU-Aware MPI}, all multi-GPU communication had to be staged through the host incurring a device $\rightarrow$ host copy on the source GPU and a host $\rightarrow$ device copy on the target GPU. Using a \textit{GPU-Aware MPI} implementation, on the other hand, a programmer can supply device buffers as parameters to the MPI call allowing communication to use the direct GPU-to-GPU data path established by GPUDirect RDMA or ROCnRDMA. In the process, \textit{GPU-awareness} eliminates redundant host $\leftrightarrow$ device copies and simplifies the communication code by eliding the need for host buffers.

MVAPICH2 was the first MPI implementation to begin actively integrating GPU-awareness into its runtime. 
Early MVAPICH2 work introduced basic GPU-awareness, transparently staging GPU-resident buffers through the host and optimizing transfers via pipelining schemes for the host $\leftrightarrow$ device and device $\leftrightarrow$ device transfers. These pipelining schemes were made possible by UVA, which allowed the library to differentiate between host and device pointers without relying on user hints. The ensuing GPU-awareness led to performance improvements over the GPU-oblivious version \cite{MVAPICH2-GPU-1, MVAPICH2-GPU-2, MVAPICH2-GPU-3}.
A follow-up work used CUDA IPC to optimize intra-node transfers, which prior had to be staged through buffers in host memory \cite{MVAPICH2-GPU-Intra-Node-CUDA-IPC}. Eventually, support was added for GPUDirect RDMA over the rendezvous protocol allowing transfers to bypass the host and eliminate redundant host $\leftrightarrow$ device copies. This reduced latencies; however, bandwidth was limited due to existing architectural limitations \cite{MVAPICH2-GDR-1}. A subsequent work added support for GPUDirect RDMA over the eager protocol rectifying the bandwidth limitation and further reducing latencies.
Additionally, a new loopback mechanism and an early version of GDRCopy \cite{GDRCopy} were used to eliminate expensive host $\leftrightarrow$ device \textit{cudaMemcpy}s \cite{MVAPICH2-GDR-2}. GDRCopy allows GPU memory to be mapped to the user address space which is optimized for small message sizes with minimal overhead \cite{GDRCopy, Designing-a-ROCm-Aware-MPI-Library-for-AMD-GPUs-Early-Experiences, Talk-GTC21-Latest-in-GPUDirect}. 
Another work extended point-to-point MPI calls to support GPUDirect Async allowing the GPU to progress the communication enqueued by the CPU, thus, optimizing the control path \cite{MVAPICH-GPUDirect-Async}. Other works have also increasingly focused on adding UVM-awareness to MVAPICH2-GDR \cite{MVAPICH-UM-Aware-1, MVAPICH-UM-Aware-2, MVAPICH-UM-Aware-3}.


\edit{Other major MPI implementations have also integrated GPU-awareness. Open MPI \cite{OpenMPI-CUDA-Aware-Support, GPU-Aware-Non-Contiguous-Data-Movement-In-Open-MPI} introduced CUDA-aware support in version 1.7.0. When using its internal backend, computation-based collectives (e.g., \texttt{MPI\_Allreduce}) may still stage data through the host \cite{gpugpuinterconnect}, a limitation avoided with UCX, which enables GPU-resident data movement and reductions. Open MPI also leverages GDRCopy \cite{GDRCopy} for small messages and CUDA IPC for intra-node communication. GPU-awareness extends to AMD devices through ROCm via UCX \cite{AMD-Lab-Notes-GPU-Aware-MPI-with-ROCm, ROCm-Documentation-GPU-Enabled-MPI, OpenMPI-ROCm-Aware-Support}. However, Khorassani et al. provide a native ROCm-aware runtime for MVAPICH2 which outperforms Open MPI with UCX on a cluster of AMD GPUs \cite{Designing-a-ROCm-Aware-MPI-Library-for-AMD-GPUs-Early-Experiences}. Open MPI has integrated the Unified Collective Communication (UCC) framework \cite{ucc}, a component of the UCX ecosystem designed to provide a unified interface for high-performance collective operations. UCC builds upon UCX’s transport layer and leverages its topology-aware mechanisms (such as awareness of NVLink, PCIe, and shared-memory hierarchies) to select efficient collective algorithms for GPU buffers. This integration enables Open MPI to automatically offload or accelerate collectives using the most suitable GPU interconnect path.

Other mainstream implementations have followed a similar evolution. MPICH, for instance, introduced GPU support in version 3.4 through its CH4 communication layer, which manages device buffers and selects efficient communication paths. Support was later extended to AMD GPUs via ROCm (since version 4.0), with Intel GPU support under development \cite{mpich-exascale}. On Cray systems, users must enable GPU support explicitly by setting \texttt{MPICH\_GPU\_SUPPORT\_ENABLED=1} \cite{Cray-MPI-CUDA-Aware-Support}. MPICH also integrates GDRCopy, GPUDirect RDMA, and IPC for optimized transfers. MPICH has adopted UCX as a network module in its CH4 device layer, and ongoing work explores the integration of UCC to enable optimized GPU collectives \cite{mpich-ucc}.
}

\subsection{GPU-Centric Collectives (GPUCCL)}
As deep learning models get ever larger, their compute requirements necessitate deploying training across multiple GPUs. Given the prevalence of collective communication in deep learning training, \edit{NVIDIA, AMD, and Intel all provide highly efficient collective communication libraries that are optimized for their respective GPU architectures. For simplicity and clarity, we refer to these vendor-specific solutions as GPU Collective Communication Libraries (GPUCCL).}
They have been integrated as a communication backend for several state-of-the-art deep learning frameworks, including Pytorch, Tensorflow, MXNet, Caffe, CNTK, and Horovod~\cite{xCCL-A-Survey-of-Industry-Led-Collective-Communication-Libraries-for-Deep-Learning}. 


The implementation of GPU-aware collectives involving computation (like reduce/allreduce) in MPI were implemented using GPU kernels for local reductions and CPU-initiated copies among the GPUs to perform aggregation. 
This approach incurs several kernel launch and communication call latencies and, additionally, requires intermediate buffers on the host. GPUCCL takes a different approach by implementing the communication and computation for the collective together in a single kernel~\cite{CUDA-Blog-Fast-Multi-GPU-collectives-with-NCCL}. Next, we highlight the differences among vendor solutions with respect to the features they support.

\subsubsection{{\bf Comparison of Collective Communication Libraries by Vendors}}

Table \ref{tab:GPUCCL2} lists the main features of the collective communication libraries supported by three main vendors: NVIDIA's NCCL, AMD's RCCL and Intel's OneCCL.

 \input{figures/GPUCCL-table}


\edit{Beyond its GPU-centric execution model, NCCL distinguishes itself through a specific philosophy of collective algorithm design. NCCL adopts a uniform design principle in which all collectives, AllReduce, AllGather, Broadcast, ReduceScatter, and AllToAll, are implemented as sequences of a small number of reusable communication primitives (send, recv, reduce, copy) applied over logical topologies determined at communicator creation. Hu et al.~\cite{hu2025demystifying} show that NCCL maps each collective onto either a ring or tree communication graph, chosen for bandwidth- or latency-dominant scenarios, and pipelines these operations using fixed-size chunks to maximize overlap between computation and communication. A key architectural element of NCCL's algorithm design is the use of parallel communication channels. Each channel corresponds to an independent instance of the collective algorithm, operating over a slice of the message. Channels allow NCCL to run multiple ring (or tree) stages in parallel, leveraging GPU thread-block parallelism and enabling bandwidth saturation across NVLink or NIC paths. As observed by Hu et al., channel count directly influences algorithmic throughput: too few channels reduce link utilization, whereas too many channels increase queueing and synchronization overheads. Channels therefore act as a structural mechanism for decomposing a collective into parallelizable sub-algorithms.

}



\edit{While NVIDIA provides NCCL, AMD offers an analogous library named \textit{RCCL (ROCm Collective Communication Library})~\cite{AMDRCCLDevGuide2025} that mirrors the NCCL API and programming abstractions, enabling compatibility for deep learning frameworks on ROCm platforms. Although the APIs are nearly identical, the performance characteristics differ significantly due to AMD's unique hardware design. RCCL operates across complex multi-die topologies, such as MI250\/MI300 accelerators, where communication traverses multiple XCD or Infinity Fabric (IF) tiers with distinct bandwidth properties. Understanding these heterogeneous bandwidth hierarchies is essential for efficient collective algorithm construction and motivates a stronger emphasis on routing and ring-selection heuristics in RCCL~\cite{10.1109/SCW63240.2024.00079}.

Beyond its GPU-centric execution model, RCCL develops a collective-algorithm design philosophy shaped by this hierarchical topology. Like NCCL, RCCL implements its core collectives by using a small set of reusable primitives. However, these primitives must be composed over a heterogeneous, multi-hop substrate, where algorithmic stages frequently cross XCD boundaries by adding non-uniform hop. As a result, RCCL constructs logical rings and trees that explicitly reflect tile-level connectivity and link asymmetries; ring segments may be lengthened, partitioned, or reordered to minimize traversal of low-bandwidth IF paths. This design makes collective execution more tightly bound to the physical GPU structure than on NVSwitch-based systems.

RCCL also adopts NCCL-style parallel communication channels to pipeline collective execution, but channel strategy is influenced by multi-tile GPU layout and ROCm partition modes (e.g., CPX/NPS4). Each channel processes a slice of the message while attempting to localize work within individual XCDs to reduce cross-tile traffic, creating a form of algorithm-level tile-aware parallelism. Hidayetoglu et al.\cite{10.1145/3650200.3656591} further observe that RCCL's latency and scaling behavior reflect both chunking strategy and the cost of multi-hop IF traversal, making channel configuration more sensitive to placement than in homogeneous NVIDIA designs.

Despite these architectural and algorithmic challenges, RCCL implements the same high-level collective patterns as NCCL—rings, trees, and hierarchical algorithms—and reuses the NCCL network-plugin ABI to support InfiniBand and RoCE transports. This feature compatibility enables shared communication backends across vendors when the underlying transport permits it. However, unlike NVIDIA-based systems equipped with NVSwitch fabrics, current AMD systems lack a switch-based all-to-all interconnect, making topology-aware collective construction and tile-level algorithm design particularly important for scaling RCCL on large multi-GPU nodes.
}

\edit{
In addition to NCCL and RCCL, Intel provides \textit{oneCCL (oneAPI Collective  Communications Library)} as part of the oneAPI ecosystem for CPU and  GPU clusters.
In contrast to NCCL/RCCL—which execute collective operations inside GPU kernels—oneCCL adopts a middleware-oriented design built over  MPI and \textit{libfabric}, delegating transport selection and topology  awareness to the underlying communication stack.
Rather than providing a purely GPU-resident collective engine, oneCCL exposes a portable set of collective communication primitives and APIs designed to operate uniformly across CPUs, integrated GPUs, and discrete Xe-series accelerators. The implementation of collectives reflects a layered orchestration model. GPU memory is managed through SYCL/Level Zero device queues, but the orchestration of collective operations is executed primarily by CPU worker threads, which schedule and advance operations on behalf of GPUs. The ATL (Abstract Transport Layer, a modular backend in oneCCL architecture, maps collective primitives onto lower-level communication fabrics such as MPI or libfabric (OFI). The ATL determines transport semantics, message progression, and endpoint selection, while Level Zero or SYCL facilitate intra-device and device–host data movement. This architecture enables oneCCL to integrate multiple accelerator types into a common collective namespace. CPU ranks and GPU ranks can participate in the same collective invocation, with oneCCL coordinating data movement across host memory, PCIe, and Xe-Link fabrics. The API ensures that collective primitives operate independently of whether the underlying buffer is allocated into CPU RAM, GPU HBM, or shared unified memory. This is particularly important on systems like Aurora, where CPUs and discrete PVC GPUs form tightly coupled multi-accelerator nodes. 

Empirical analyses by Kwack et al.~\cite{11078548} show that oneCCL unifies collective communication across these diverse endpoints, with transport-level behaviors delegated to MPI/libfabric on Slingshot networks. However, oneCCL’s abstraction comes with implications for communication progress and GPU-centric collective design. Unlike NCCL and RCCL, whose collective kernels run entirely on GPUs, oneCCL’s GPU participation remains dependent and host-driven. Therefore, collective primitives operate through command queues and event dependencies, but Level Zero kernels do not independently initiate network activity. As a result, overlap and latency characteristics depend on CPU worker scheduling and the performance of the ATL backend.

This distinction is highlighted by recent GPU-aware MPI studies~\cite{10171511, 10.1145/3626203.3670549}, which demonstrate that direct Level Zero and IPC-based approaches can outperform oneCCL for GPU–GPU collectives when host intervention becomes a limiting factor. 
Performance analysis study~\cite{10.1145/3650200.3656591} further shows that oneCCL exhibits different scaling patterns than NCCL/RCCL on hierarchical multi-GPU systems precisely because its collective primitives are executed through a hybrid CPU/GPU control structure.
}

\subsubsection{\textbf{Further Work on Collectives}} 
\label{sec:further-work-on-collectives}
\edit{
Despite its significant benefits, GPUCCL presents several inherent challenges, primarily stemming from resource contention and its static abstraction model. 
First, a fundamental issue of using GPU threads for communication \textit{and} computation is that the two routines now contend for the same limited resource. In this case, if computation is scheduled ahead of the collective, it can monopolize all GPU resources, effectively serializing the collective behind the computation. One workaround around this is to launch the collective on a higher-priority stream such that it is always scheduled first \cite{Benchmarking-multi-GPU-Applications-on-Modern-multi-GPU-integrated-systems, CUDA-Blog-Fast-Multi-GPU-collectives-with-NCCL}. 
Moreover, GPUCCL's execution model is tailored for regular collective patterns where communication size and data type are statically expressed as host arguments. While suitable for deep learning, this design lacks the flexibility needed for more dynamic communication scenarios \cite{ncclx}. Furthermore, NCCL's design often restricts interconnects to a single transfer mode (such as thread-copy over DMA-copy), and its conservative synchronization methods hinder the implementation of advanced optimization techniques, such as fine-grained double-buffering to effectively hide communication latency \cite{msccl}.}

\edit{To address these aforementioned issues, several collective communication libraries have been proposed; notably, many of the solutions developed by industry partners are now open-sourced and readily available for use.}

 
 Blink is a collective communication library with the express goal of achieving optimal link utilization. To do this, Blink detects the underlying topology, models the topology as a graph, and then uses a technique known as \textit{packing spanning trees} to dynamically generate communication primitives. As a result, Blink was shown to reduce model training time on an image classification task by 40\% compared to NCCL2 \cite{Blink-Fast-and-Generic-Collectives-for-Distributed-ML}. 
 
Dryden et al. present Aluminum, a GPU-aware library for large-scale training of deep neural networks. Aluminum extends NCCL with tree-based algorithms to avoid the latency bottlenecks of its default ring implementation. Additionally, it adds support for non-blocking NCCL allreduce operations. For MPI, Aluminum gets around forced synchronizations stemming from the MPI-GPU semantic mismatch by associating a single GPU stream with an MPI communicator and synchronizing with respect only to that stream. The resulting optimizations bring about speedups compared to GPU-aware MPI and NCCL-based implementations \cite{Aluminum-An-Asynchronous-GPU-Aware-Communication-Library-Optimized-for-Large-Scale-Training-of-Deep-Neural-Networks-on-HPC-Systems}.


\edit{
Several recent research systems aim to extend or specialize the capabilities of vendor-provided GPUCCL libraries. Among these, NCCLX~\cite{ncclx}, developed by Meta, introduces a transparent acceleration layer for NCCL specifically targeting deep-learning workloads. NCCLX observes that most data-parallel training jobs repeatedly invoke the same collective patterns which are predominantly AllReduce and AllGather—over fixed message shapes. By exploiting this regularity, NCCLX constructs optimized execution paths inside NCCL’s existing kernels through collective fusion, path coalescing, and specialized kernel pipelines that eliminate redundant synchronization and memory movement stages. Crucially, NCCLX preserves NCCL’s public API and requires no modification to frameworks such as PyTorch or TensorFlow, making it a performance enhancement for standard deep-learning training stack. 

Unified Collective Communication (UCC) \cite{ucc} is an open-source project providing an API and library implementation of high-performance and scalable collective operations, leveraging topology-aware algorithms and techniques such as in-network computing and DPU offloading. It relies on UCX point-to-point communication, as well as on NCCL/RCCL, SHARP, and others. UCC offers a portable, backend-agnostic framework that unifies collectives across CPUs, GPUs, and DPUs by dynamically selecting among UCX-based implementations,  SHARP offload, and vendor libraries such as NCCL and RCCL. Similarly, Microsoft Collective Communication Library (MSCCL)~\cite{msccl}, introduces a low-level GPU-centric communication substrate built around device-resident primitives (put, signal, wait, flush) and a high-level DSL for algorithm synthesis, enabling fine-grained, fused communication–computation kernels that outperform traditional NCCL paths on emerging AI workloads such as LLMs. MSCCL++ outperforms NCCL and MSCCL by up to 2.8x and 1.6x for small messages, and by up to 2.4x and 2.0x for large messages, respectively. Hierarchical Collective Communication Library (HiCCL)~\cite{hiccl} adopts a hierarchy-aware design that decomposes collective operations into universal primitives (multicast, reduction, fence) and maps them onto multi-tier accelerator topologies—including GPU tiles, multi-GPU nodes, and multi-NIC clusters—achieving portable performance across NVIDIA, AMD, and Intel systems. Many other efforts are going in the direction of unifying the offloading collective operations into the MPI runtime~\cite{mpixccl}.

Finally, MPI-xCCL~\cite{mpixccl} integrates NCCL, RCCL, and HiCCL to create a hybrid execution model in which MPI transparently offloads collective operations to vendor libraries when beneficial, while retaining MPI semantics and enabling collectives not supported natively by GPUCCLs.


Beyond portability and topology-awareness, recent research has also begun to explore collective communication optimization along new axes such as energy efficiency and configuration auto-tuning. PCCL~\cite{PCCL} introduces a power-aware collective communication layer built on top of NCCL that exploits the observation that many collective kernels, particularly \textit{AllReduce} and \textit{AllGather} operations in LLM workloads, are frequency-insensitive and can run at substantially reduced GPU clock frequencies without measurable bandwidth loss. By integrating fine-grained DVFS management into the collective call path, PCCL identifies the minimum safe GPU frequency for each collective, and automatically inserts frequency-set and frequency-reset events around NCCL kernels. 

Complementary to power-centric optimization, many libries address the challenge of optimising a large configuration space (algorithm, protocol, transport, channels, threads, chunk sizes) that the default GPUCCL cost model can be sub-optimal for many message sizes, GPU topologies, and communication patterns~\cite{TuCCL,autoccl}. AutoCCL~\cite{autoccl} provides an automated, online tuning framework that profiles collective execution during the early iterations of training, classifies parameters into implementation-level versus resource-allocation parameters.
Unlike prior offline tuners, AutoCCL directly accounts for communication–computation balancing, frequently achieving 1.2–1.8x bandwidth improvements over NCCL on both PCIe and NVLink systems and up to 32\% end-to-end iteration-time improvements for LLM workloads. The approach highlights the growing need for collective runtimes that dynamically adapt their execution strategy to hardware conditions, workload patterns, and runtime interference.}




\subsection{GPU-centric OpenSHMEM (GPUSHMEM)}
\edit{
GPU-centric OpenSHMEM runtimes are NVIDIA's NVSHMEM\cite{NVSHMEM}, AMD's ROC\_SHMEM\cite{ROCSHMEM} and Intel SHMEM\cite{IntelShmem} libraries. 
Given its earlier inception, we first discuss NVSHMEM and, in doing so, introduce concepts fundamental to all three libraries. 
Note that NVSHMEM and ROC\_SHMEM are GPU-centric PGAS runtimes designed from scratch for their respective GPU ecosystems. Both libraries follow the OpenSHMEM programming model, but they introduce their own runtime structures and calling conventions, lack the context abstraction present in the OpenSHMEM specification, and expose host APIs that operate only on device-resident symmetric memory. In contrast, Intel SHMEM adheres closely to the OpenSHMEM specification and supports both host and device pointers via a unified SYCL-based API.}

\subsubsection{\textbf{NVSHMEM}} \label{NVSHMEM}
NVSHMEM is NVIDIA's implementation of the OpenSHMEM specification for CUDA devices. NVSHMEM is a Partitioned Global Address Space (PGAS) library that provides efficient one-sided \textit{put / get} APIs for processes to access remote data objects. NVSHMEM supports  point-to-point and collective communication between GPUs both within and across nodes \cite{NVSHMEM}.  

NVSHMEM works on the concept of a \textit{symmetric heap}. During NVSHMEM initialization, each process that is mapped to a GPU, referred to as a processing element (PE) reserves a block of GPU memory using \textit{nvshmem\_malloc()}. In NVSHMEM, all memory allocations must be performed collectively, meaning that all symmetric memory regions within the heap must have identical sizes and must be allocated at the same time. To access remote memory on a different PE, a given PE requires the offset for the symmetric memory as well as the rank of the remote PE.

In addition, NVSHMEM provides APIs for synchronizing a group of PEs. These APIs comprise signal-wait mechanisms that can serve as a means for point-to-point synchronization and collective synchronization calls that can function as global barriers. This feature is particularly important since there is a general lack of kernel-side global barriers, with the CPU typically performing the role of global synchronizer for devices. The capacity for a device to synchronize efficiently across the device without terminating kernel execution is a crucial prerequisite for transferring the control plane to the GPU.

A notable attribute of NVSHMEM is that it offers both host-side and device-side APIs. The host-side APIs expose an optional stream argument that can be used to implement communication-computation overlap. For certain calls, the GPU-side variants provide the calls in three granularities: thread, thread block, and warp. The thread variant means that the call should be performed by a single device thread and will be executed by that thread. The thread block and warp variants use multiple threads to execute the communication call cooperatively. These variants should be called by all threads in the corresponding thread block or warp. A previous performance comparison between the host-side and device-side APIs found negligible differences in performance between the two, with host-side APIs slightly outperforming the device-side variants \cite{Performance-Trade-offs-in-GPU-Communication-A-Study-of-Host-and-Device-Initiated-Approaches}. This study was conducted using an early version of NVSHMEM (0.3.0), which has since seen improvements in GPU-side API performance.

As of version 2.7.0, NVSHMEM introduced the Infiniband GPUDirect Async (IBGDA) transport built on top of GPUDirect Async \cite{NVSHMEM-2.7.0-Release-Notes}. The IBGDA transport allows GPUs to issue inter-node communication directly to the NIC, bypassing the CPU entirely. Without IBGDA, device-side inter-node communication calls are performed through a proxy thread on the CPU that triggers the corresponding NIC operations. This proxy thread consumes CPU resources and creates a bottleneck in achieving peak NIC throughput for fine-grained transfers \cite{Blog-Improving-Network-Performance-of-HPC-Systems-Using-NVIDIA-Magnum-IO-NVSHMEM-and-GPUDirect-Async}. NVSHMEM with IBGDA support, combined with persistent kernels, enables the complete transfer of both data and control paths to the GPU and marks a significant shift towards fully autonomous multi-GPU execution. However, as discussed in Section \ref{GPUDirect RDMA}, GPUDirect RDMA only enforces GPU-NIC memory consistency across kernel boundaries. 
This inherent reliance on the CPU for memory consistency is a potential obstacle toward truly autonomous multi-GPU execution. One workaround is using a callback mechanism whereby the persistent kernel signals the CPU to perform a consistency-enforcing API call (i.e., \textit{cudaDeviceFlushGPUDirectRDMAWrites()}). The efficacy of this solution is unclear and warrants further investigation. Enforcing GPU-NIC memory from \textit{inside the kernel} is supported by ROC\_SHMEM, which we discuss in the next section.


In recent years, NVSHMEM has been integrated as a communication backend into multiple runtimes. PETSc implemented PetscSF, a scalable communication layer based on NVSHMEM, to complement their MPI-based approach, which did not work well with CUDA stream semantics and prevented kernel launch pipelining \cite{PetscSF-Scalable-Communication-Layer}. Kokkos Remote Spaces, adding distributed memory support to the Kokkos programming model, uses NVSHMEM as a communication backend \cite{Kokkos-Remote-Spaces-Repository, Talk-Early-Experience-with-NVSHMEM-Extending-the-Kokkos-Programming-Model-with-PGAS-Semantics}. An NVSHMEM implementation of the Kokkos Conjugate Gradient Solver outperforms the CUDA-aware MPI implementation while significantly reducing the code base size \cite{CUDA-Blog-Scaling-Scientific-Computing-with-NVSHMEM}. Choi et al. use persistent kernels and NVSHMEM to implement CharminG, a fully GPU-resident runtime system inspired by Charm++ \cite{CharminG-A-Scalable-GPU-Resident-Runtime-System}. The Livermore Big Artificial Neural Network (LBANN) implements a spatial-parallel convolution using NVSHMEM that outperforms MPI and Aluminium implementations \cite{CUDA-Blog-Scaling-Scientific-Computing-with-NVSHMEM}. QUDA, a library for lattice QCD computations, has used NVSHMEM and persistent kernels for improved strong scaling of Dirac operators \cite{QUDA-Repository, Talk-NVSHMEM-Overcoming-Latency-Barriers}.

NVSHMEM has also been used outside of runtime-based approaches to achieve performance improvements. Chu et al. combine NVSHMEM with persistent kernels to implement a state-of-the-art GPU-based key-value store \cite{Designing-High-Performance-In-Memory-Key-Value-Operations-with-Persistent-GPU-Kernels-and-OpenSHMEM}. Xie et al. use NVSHMEM to implement a single-node multi-GPU sparse triangular solver (SpTRSV) that achieves good performance scalability compared to a UVM-based design \cite{Fast-and-Scalable-Sparse-Triangular-Solver-for-Multi-GPU-Based-HPC-Architectures}. Ding et al. combine persistent kernels with NVSHMEM for impressive performance in a sparse triangular solver (SpTRSV) on single- and multi-node setups. Atos implements both persistent and discrete kernels with NVSHMEM-based communication to achieve state-of-the-art performance on multi-GPU BFS within and across nodes \cite{Atos-Vyse-Scalable-Irregular-Parallelism-with-GPUs-Getting-CPUs-Out-of-the-Way}. Wang et al. propose MGG, a system design that accelerates Graph Neural Networks (GNNs) on multi-GPU systems using a GPU-centric software communication-computation pipeline with NVSHMEM for fine-grained communication \cite{MGG-Accelerating-Graph-Neural-Network-with-Fine-grained-intra-kernel-Communication-Computation-Pipelining-on-Multi-GPU-Platforms}. Ismayilov et al. use persistent kernels and device-side NVSHMEM to implement fully GPU-side Jacobi 2D/3D and CG solvers that outperform CPU-controlled baselines. They reserve some thread blocks for communication while others handle computation, a technique they call \textit{thread block specialization}, to achieve \textit{explicit} device-side communication-computation overlap \cite{Multi-GPU-Communication-Schemes-for-Iterative-Solvers-When-CPUs-Are-Not-in-Charge}. Punniyamurthy et al. use ROC\_SHMEM and persistent kernels to overlap embedding operations with collective communication in deep learning recommendation models \cite{GPU-initiated-Fine-grained-Overlap-of-Collective-Communication-with-Computation}.


\subsubsection{\textbf{ROC\_SHMEM}}
ROC\_SHMEM is AMD's implementation of the OpenSHMEM specification for AMD GPUs. ROC\_SHMEM offers two communication backends. The first, known as GPU-IB, implements Infiniband on the GPU, similar to NVSHMEM's IBGDA transport. The second, called Reverse Offload (RO), uses host-side proxy threads and offloads communication to the CPU. GPU-IB is the default backend and offers the best performance \cite{ROCSHMEM}. ROC\_SHMEM works almost identically to NVSHMEM and offers analogous APIs. However, there are several significant differences.
First, as mentioned in the previous section, NVSHMEM runs into GPU-NIC memory consistency problems when intra-node communication is issued from persistent kernels. ROC\_SHMEM, on the other hand, explicitly addresses this issue and guarantees correctness when persistent kernels are being used.  Hamidouche et al. analyze the GPU-NIC memory mismatch stemming from the GPU's relaxed memory model and propose changes integrated into ROC\_SHMEM \cite{GIO-GPU-Initiated-OpenSHMEM-Correct-and-Efficient-Intra-Kernel-Networking-for-dGPUs}. This means that ROC\_SHMEM provides completely CPU-free communication mechanism that can move the entire flow of multi-GPU execution to the device. 


Second, ROC\_SHMEM uses GPU \textit{shared memory} (\textit{local data store (LDS)} in AMD parlance) to store network state for faster access. To the best of our knowledge, this optimization is not implemented in NVSHMEM. While this is most likely beneficial for execution time, the increased shared memory message could limit occupancy and negatively impact performance \cite{GPU-initiated-Fine-grained-Overlap-of-Collective-Communication-with-Computation, AMD-Instinct-MI200-Instruction-Set-Architecture}. 

Third, prior versions of ROC\_SHMEM required allocating symmetric buffers as uncacheable in order to prevent stale data from being communicated. However, as AMD recently introduced intra-kernel cache flush instructions, the data can be flushed before initiating the network transaction, allowing the data to be cached. No such instructions are provided by NVIDIA, meaning that NVSHMEM buffers are likely allocated as uncacheable \cite{GPU-initiated-Fine-grained-Overlap-of-Collective-Communication-with-Computation}.


\subsubsection{\textbf{Intel SHMEM}}
\edit{Intel recently introduced Intel SHMEM \cite{IntelShmem}, the first GPU-aware implementation of OpenSHMEM for Intel GPUs. It allows SHMEM routines to operate directly on GPU memory and supports GPU-initiated communication by embedding SHMEM calls inside SYCL kernels. 
Intel SHMEM integrates with the SYCL programming model, offering a portable C++ interface for heterogeneous systems. In contrast, existing solutions such as NVSHMEM and ROC\_SHMEM provide similar capabilities but are tied to vendor-specific ecosystems. 

To maintain compatibility with the OpenSHMEM 1.5 specification, Intel SHMEM supports both device- and host-side APIs for point-to-point operations, collective operations via the teams API, and SYCL-specific extensions for work-group and sub-group communication. Inter-node communication is handled through a host proxy thread, which forwards GPU-initiated operations to a standard OpenSHMEM backend. In other words, 
GPU issues SHMEM operations inside a SYCL kernel. The GPU writes work queue entries into a memory region accessible by both GPU and host.
The host thread running on CPU detects the GPU-initiated RMA operation. The host executes actual RDMA using a standard OpenSHMEM library.
The current implementation relies on Sandia OpenSHMEM (SOS), leveraging its robust support for OFI/libfabric transports and its capability to maintain a symmetric heap in GPU memory. 
This layered design enables Intel SHMEM to provide functional equivalence with NVSHMEM and ROC\_SHMEM while integrating cleanly with Intel’s broader oneAPI and SYCL ecosystem.

For intra-node communication, Intel SHMEM can perform GPU-to-GPU data movement directly without involving the host CPU.
This is achieved by allowing the GPU to issue SHMEM operations from within SYCL kernels, using Intel’s Level Zero–based backend to translate these operations into: (1)
direct GPU load/store transfers, when GPUs share a unified memory fabric, or
(2) GPU copy-engine transfers, which bypass the host CPU and use dedicated DMA engines on the GPU.


}

\subsection{Comparison and Discussion of User-Level Libraries}

\edit{
While GPU-aware MPI, GPUCCL, and GPUSHMEM  all offer mechanisms for programming multi-GPU systems, significant differences exist in their semantics and performance characteristics. Key distinctions include stream support, API location, programming and performance.

\subsubsection{\bf Streaming Support}
\label{sec:streams}
GPUs operate on the concept of streams which are command queues that guarantee ordering among GPU operations. The GPU scheduler ensures that kernels and other operations launched on a stream execute in the order they were enqueued and do so with correct data dependencies. Since kernel launches are asynchronous and do not block the host, GPU runtimes can pipeline kernel launches and overlap the launch latencies behind kernel execution. The \textit{semantic mismatch} between the MPI and GPU models is that MPI has no awareness of GPU streams. As a result, it is not possible to enqueue an MPI call on a given GPU stream or for a GPU stream to wait on the completion of a pending MPI routine. By implication, interlacing MPI calls with GPU kernels will require host-blocking synchronizations in order to maintain data correctness. For example, before initiating an MPI send, the programmer has to block the host to synchronize all streams which operate on the send buffer. Similarly, waiting on completion of pending MPI communication will also require host-blocking synchronization. In either case, these forced synchronizations impair kernel launch pipelining, prevent opportunities for overlap and force the programmer into alternating bulk phases of communication and computation \cite{Aluminum-An-Asynchronous-GPU-Aware-Communication-Library-Optimized-for-Large-Scale-Training-of-Deep-Neural-Networks-on-HPC-Systems, PetscSF-Scalable-Communication-Layer, CPU-and-GPU-initiated-Communication-Strategies-for-CG, uniconn}.

We see two possible non-mutually exclusive paths for resolving the semantic mismatch. The first is making MPI runtimes stream-aware by adding an explicit stream parameter to MPI routines. This would solve the issue of impaired kernel launch pipelining and allow MPI calls to seamlessly integrate into GPU runtimes. The second is providing the option of device-initiated MPI calls. This would reduce the programmer burden of juggling two distinct programming models and, additionally, provide implicit communication-computation overlap. Both directions have been explored in the literature on a limited scale. The FLAT compiler automatically converts device-side MPI calls to their host-side equivalents \cite{Flat-A-GPU-Programming-Framework-to-Provide-Embedded-MPI}. dCUDA implements device-side operations with MPI semantics but uses CPU helper threads for the actual communication. They rely on the GPU's inherent memory latency hiding capabilities to \textit{implicitly} overlap communication with computation, ultimately outperforming a GPU-aware MPI baseline \cite{dCUDA-Hardware-Supported-Overlap-of-Computation-and-Communication}. Namashivayam et al. explore new communication schemes to introduce GPU stream-awareness in MPI. They use the \textit{triggered operations} feature on HPE Slingshot 11 interconnect, allowing the CPU to enqueue communication and synchronization operations to the NIC, which the GPU can then trigger. This reduces CPU involvement in the critical path and eliminates expensive synchronizations. While inter-node experiments show some performance benefits, the proposed scheme falters in intra-node setups as progress threads need to be used to emulate deferred execution semantics \cite{Exploring-GPU-Stream-Aware-Message-Passing-using-Triggered-Operations}. A follow-up work eliminates progress threads for intra-node communication, opting to use P2P Direct Load Store-based GPU kernels and GPU IPC-based mechanisms instead. The evaluation shows performance improvements over stream-oblivious MPI baselines \cite{Exploring-Fully-Offloaded-GPU-Stream-Aware-Message-Passing}.

More recently, MPIX streams~\cite{10.1145/3555819.3555820} allow an application to explicitly map its GPU execution stream context and pass it to the MPI library. This enables the MPI implementation to operate directly on GPU streams, thereby eliminating unnecessary synchronization overhead and improving the performance of GPU-to-GPU communication.
However, none of these solutions have yet been incorporated into the official MPI standard.
}

\subsubsection{\textbf{Host vs Device-side API}}
\edit{Both GPU-aware MPI and GPUCCL utilize host-side APIs, mandating that communication routines and their parameters (like size and data type) are  expressed on the host CPU using host arguments. GPUSHMEM offers both host and device-side APIs, enabling communication to be invoked directly from a GPU device kernel with input arguments defined as device variables. This device-centric approach allows GPUSHMEM to better handle low-latency and dynamic communication patterns. Data transfers are initiated within the computation kernel and sent straight to the network, minimizing latency and making the approach ideal for fine-grained communication.

However, this device-side invocation introduces programming and performance measurement challenges. Mixing computation and communication within a single kernel complicates the programming model, and since current profiling tools only operate at the kernel granularity, performance profiling  is non-trivial. Moreover, especially when combined with persistent kernels, the technique causes resource contention as streaming processors and threads must be shared between the two. When employing techniques like dividing thread blocks (TBs) for overlap, synchronizing the dedicated communication TBs versus the computation TBs becomes difficult. While Thread Block Clusters (TBCs) can potentially mitigate this synchronization issue, their programming further complicates the overall programming complexity.

Mimicking NVSHMEM, NVIDIA NCCL 2.28 \cite{NCCL-2.28-release-notes} introduced a new device-side API to enable fusing communication and compute. This was a significant departure from earlier versions where all NCCL operations were host-initiated. The new API allows GPU kernels to directly initiate data movement, and its use requires setting up data buffers with symmetric memory windows to facilitate direct GPU-to-GPU communication.

}

\input{listings/mpi-example}

\subsubsection{\textbf{Programming Examples}}
\edit{
Listings \ref{lst:mpi-example}, \ref{lst:nccl-example}, \ref{lst:nvshmem-example} present simplified one-way bandwidth benchmarks using GPU-aware MPI, NCCL, and device-side NVSHMEM. These examples highlight the semantic differences among the three communication libraries. While all three take a buffer pointer and size, MPI and NCCL rely on a two-sided communication model with synchronous send/receive semantics. In contrast, NVSHMEM employs a one-sided model: put/get operations are asynchronous with respect to the remote GPU. This distinction is visible in Listings \ref{lst:mpi-example} and \ref{lst:nccl-example} versus Listing \ref{lst:nvshmem-example}, where NVSHMEM requires the sender to specify the receiver’s buffer address directly.}

\input{listings/nccl-example}
\input{listings/nvshmem-example}
  \edit{
MPI does not expose a stream parameter as discussed in Section \ref{sec:streams}, whereas NCCL and NVSHMEM operations are explicitly tied to a CUDA stream. NCCL also supports grouping multiple operations between groupStart/groupEnd to amortize launch overhead. Because MPI lacks stream semantics, MPI programs must explicitly synchronize CPU and GPU progress e.g., using {\em Waitall} to ensure completion. 
NVSHMEM kernels that invoke device-side NVSHMEM synchronization or collective APIs
(e.g., \texttt{nvshmem\_wait}, \texttt{nvshmem\_barrier}, or collective operations)
must be launched with \texttt{nvshmemx\_collective\_launch}. Otherwise, behavior is undefined. Because this mechanism relies on CUDA cooperative launch, the grid size is constrained so that the participating thread blocks can execute concurrently on the GPU.
If a kernel uses only one-sided put/get operations and does not use device-side NVSHMEM synchronization or collective operations, a regular kernel launch is permitted.


Supporting multiple communication libraries often forces developers to reimplement communication backends, reducing productivity. To address this, Sagbili et al. proposed a unified communication interface that consolidates these models under a single API and demonstrated performance comparable to naive implementations \cite{uniconn}.
}

\subsubsection{\textbf{Performance}}
\edit{Several works compare the performance of different user-level libraries, either on real applications or through microbenchmarks. The authors of \cite{CPU-and-GPU-initiated-Communication-Strategies-for-CG} implemented standard and pipelined Conjugate Gradient (CG) using MPI, NCCL/RCCL, and NVSHMEM on both NVIDIA and AMD GPUs. They found that avoiding CPU-GPU synchronization via streams improved CG performance significantly by 5–15\%. For NVIDIA systems, they suggested using NCCL for small-message AllReduce but using MPI for point-to-point. On AMD GPUs, MPI outperformed vendor solutions, which was attributed to the less mature vendor software.

The benefits of NVSHMEM’s GPU-initiated communication have been demonstrated in molecular dynamics simulations using GROMACS \cite{Doijade_2025}, yielding performance improvements of up to 2x over MPI. These gains are attributed to the suitability of the PGAS model, which removes the CPU from the critical path and enables implementation-level optimizations such as kernel fusion. However, with the recent introduction of the NCCL \textit{Device API}, new studies suggest that its performance is now comparable to NVSHMEM for both point-to-point and all-to-all communication.

Similar advantages have been observed in graph processing applications, specifically Breadth-First Search (BFS)~\cite{10.1007/978-3-319-73814-7_6}. In these workloads, the inherent irregular and fine-grained access patterns align more effectively with the NVSHMEM programming model than with standard MPI. Similar findings have been reported for other irregular workloads such as updates of distributed Bloom filters and the search of connected components in large-scale graphs~\cite{10.1145/3708035.3736073}. 

Other works~\cite{gpugpuinterconnect} compare NCCL (and RCCL) to GPU-Aware MPI on both collective and point-to-point operations on three supercomputers. The results show a tendency for MPI to perform better in point-to-point operations, while *CCL performs better in collectives. However, this also depends on the specific system and optimizations provided by those libraries, and sometimes the best solution depends both on the number of nodes and vector size. Moreover, the paper reports instability in RCCL at high node count, as also highlighted in a comparison between Cray MPICH and RCCL performance on the Frontier supercomputer~\cite{singh2025bigsendoffhighperformance}. 

Another limitation reported for both Open MPI~\cite{gpugpuinterconnect} and Cray MPICH~\cite{singh2025bigsendoffhighperformance} concerns the execution of reduction-based collectives, such as \texttt{MPI\_Allreduce} and \texttt{MPI\_Reduce\_scatter}. Unless specifically configured with accelerator-offload components, these MPI implementations often default to a host-staging protocol: data is copied to the host, reduced using the CPU, and then copied back to the device. This round-trip data movement and reliance on CPU arithmetic capabilities significantly degrade performance compared to native GPU collective libraries, which perform reductions directly on the GPU. This was also addressed in a study optimizing the MFDn nuclear configuration interaction code \cite{11018300}. The paper demonstrated that replacing the baseline MPI/\texttt{OpenACC} implementation with native CUDA kernels for computation and high-performance communication protocols yielded substantial gains. For large, network-bandwidth-bound problem sizes, the NCCL/CUDA approach proved most effective, achieving up to a 4.9x speedup over the baseline by eliminating the host-staging for reductions and fully leveraging GPU-native communication and parallelism.

Despite these variations in programming and performance, all three libraries are mature, with diverse implementations available; however, new GPUCCL variants as discussed in Section \ref{sec:further-work-on-collectives} are actively emerging from industry and research to address evolving AI application needs.
}

\subsubsection{\textbf{Library Interactions}}
\begin{figure}
    \centering
    \includegraphics[width=.7\linewidth]{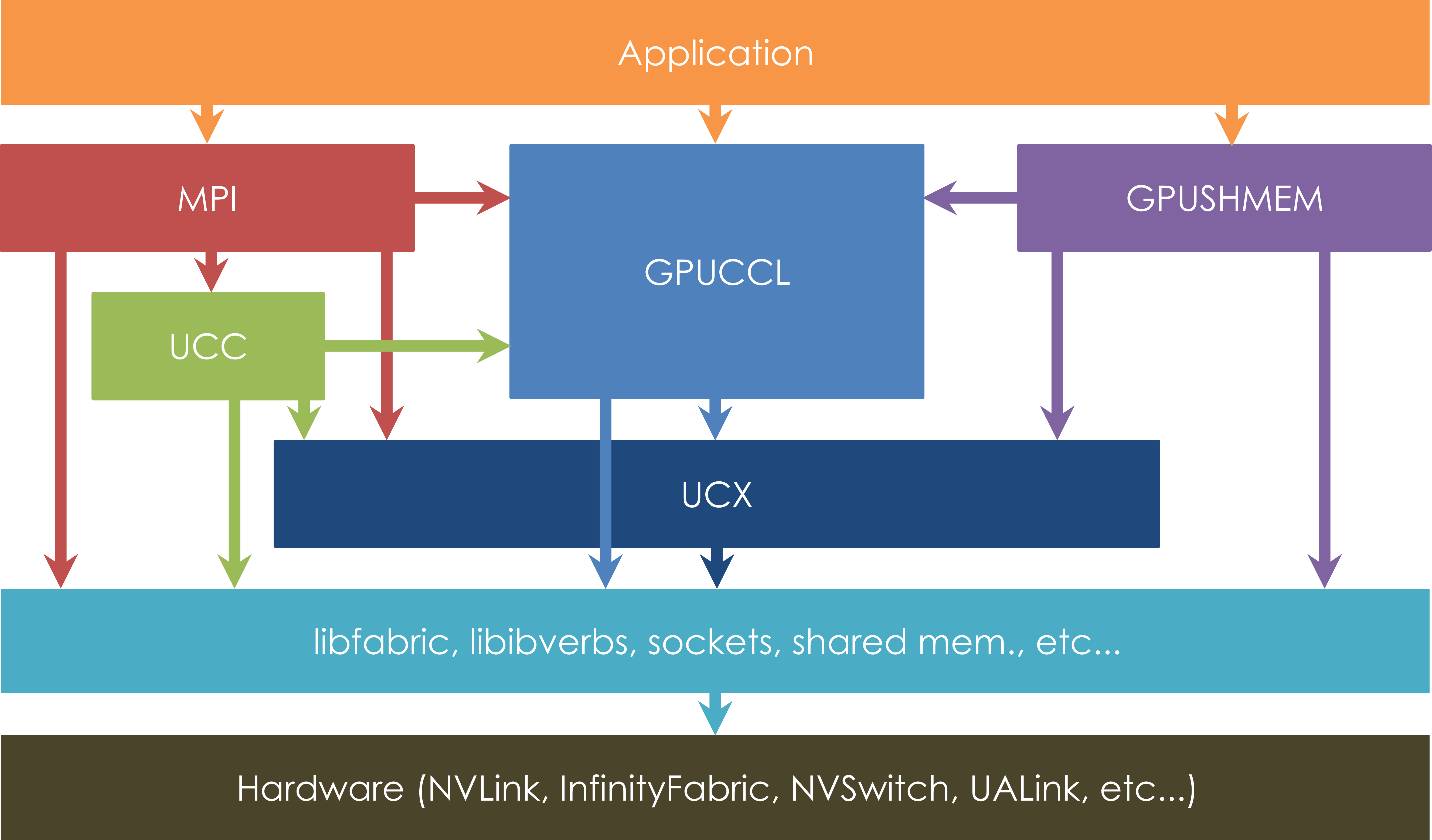}
    \caption{Interactions between different GPU-GPU communication libraries. Figure is available under CC-BY \cite{figure-cc-by}.}
    \label{fig:swstack}
\end{figure}

\edit{Figure~\ref{fig:swstack} illustrates the software stack hierarchy from the application down to the hardware. Applications typically interface directly with high-level libraries such as MPI, GPUSHMEM, or GPUCCL. Lower-level communication libraries, such as UCX or libfabric, generally serve as middleware used by these high-level frameworks rather than being accessed directly by applications. While legacy MPI implementations often ran directly over transport interfaces (e.g., libfabric or libibverbs), modern GPU-aware MPI stacks are more modular. They typically layer on top of UCX for point-to-point operations and UCC for collectives. In some cases, such as with MVAPICH, MPI may also offload collective operations directly to GPUCCL to leverage its topology-aware optimizations.

Similarly, GPUSHMEM adopts a hybrid architecture: it utilizes UCX or direct transport interfaces (libfabric/libibverbs) for low-latency point-to-point operations, while layering on top of GPUCCL to accelerate collective routines. Finally, GPUCCL typically interfaces directly with the transport layer (libfabric or libibverbs), though it can also be configured to run on top of UCX via specific plugins for enhanced portability. Last, UCC serves as a higher-level abstraction, executing collective operations by either offloading them to GPUCCL or employing its own implementations over UCX and low-level transports.}

%% file: figures/GPUCCL-table.tex
\begin{table}[t]
\centering
\small
\edit{
\caption{Comparison of vendor collective communication libraries in terms of collective primitives, execution model, accelerator integration, and transport semantics. 
Abbreviations: B=broadcast, R=reduce, AR=allreduce, RS=reducescatter, AG/AGv=allgather and variants, AA=all-to-all, P2P=point-to-point.}
\label{tab:GPUCCL2}
\setlength{\tabcolsep}{3pt}
\begin{tabular}{l p{0.24\columnwidth} p{0.24\columnwidth} p{0.24\columnwidth}}
\toprule
\textbf{Library} & \textbf{NCCL} & \textbf{RCCL} & \textbf{oneCCL} \\
\midrule
\midrule

\textbf{Collective primitives} &
AR, R, B, AG/AGv, RS, AA, \quad P2P &
AR, R, B, AG/AGv, RS, AA,\quad \quad P2P &
AR, R, B, AG/AGv, RS, AA/AAv, \quad P2P (async) \\
\midrule
\textbf{API model} &
CUDA streams; GPU-kernel-driven &
HIP streams;  GPU-kernel-driven &
Unified SYCL/L0 API across CPU+GPU ranks \\
\midrule
\textbf{Execution engine} &
GPU kernels (device-autonomous) &
GPU kernels (tile-aware) &
CPU workers + device queues \\
\midrule
\textbf{Accelerator integration} &
NVIDIA GPUs only; CPU proxy threads &
AMD MI-series multi-tile GPUs; explicit locality &
CPUs + Intel PVC GPUs; unified collective namespace \\
\midrule
\textbf{Intra-node fabric} &
NVLink / NVSwitch &
xGMI / Infinity Fabric &
Xe-Link + shared memory \\
\midrule
\textbf{Transport backend} &
NCCL-NET (IB/RoCE, TCP); GPU-triggered NIC operations &
NCCL-NET plugins; HIP-aware routing &
ATL with MPI or libfabric; SYCL/L0 transport \\
\midrule
\textbf{Progress \& initiation} &
GPU-initiated NIC operations &
GPU-initiated operations; tile-level routing &
CPU-initiated network operations; GPU via SYCL/L0 \\
\midrule
\textbf{Topology awareness} &
Automatic NVLink/NVSwitch graph construction &
XCD topology; hop-aware \quad routing &
Provided by MPI/libfabric \\
\midrule
\textbf{Distinctive property} &
Fully GPU-autonomous communication &
GPU-driven, topology-aware collectives &
Heterogeneous, transport-modular model \\

\midrule
\bottomrule
\end{tabular}
}
\end{table}

%% file: listings/mpi-example.tex
\begin{figure*}[htb]
\centering
\begin{lstlisting}[caption={Simple one-way bandwidth benchmark with GPU-aware MPI}, label={lst:mpi-example}]
//Host-side API 
if (rank == 0) { // process 0
    for (int j = 0; j < window_size; ++j) {
        MPI_Isend(send_buf, message_size, MPI_FLOAT, 1, 100, MPI_COMM_WORLD, send_request + j);
    }
    MPI_Waitall(window_size, send_request, reqstat);
    MPI_Recv(recv_buf, 1, MPI_FLOAT, 1, 101, MPI_COMM_WORLD, reqstat);
} else {  // process 1
    for (int j = 0; j < window_size; ++j) {
        MPI_Irecv(recv_buf, message_size, MPI_FLOAT, 0, 100, MPI_COMM_WORLD, recv_request + j);
    }
    MPI_Waitall(window_size, recv_request, reqstat);
    MPI_Send(send_buf, 1, MPI_FLOAT, 0, 101, MPI_COMM_WORLD);
}
    \end{lstlisting}
\end{figure*}

%% file: listings/nccl-example.tex
\begin{figure*}[htb]
\centering
    \begin{lstlisting}[caption=Simple one-way bandwidth benchmark with NCCL, label={lst:nccl-example}]
//Host-side API
if (rank == 0) { // process 0
    ncclGroupStart();
    for (int j = 0; j < window_size; ++j) {
        ncclSend(send_buf, message_size, ncclFloat, 1, comm, stream);
    }
    ncclGroupEnd();
    ncclRecv(recv_buf, 1, ncclFloat, 1, comm, stream);
} else {  // process 1
    ncclGroupStart();
    for (int j = 0; j < window_size; ++j) {
        ncclRecv(recv_buf, message_size, ncclFloat, 0, comm, stream);
    }
    ncclGroupEnd();
    ncclSend(send_buf, 1, ncclFloat, 0, comm, stream);
}
    \end{lstlisting}
\end{figure*}

%% file: listings/nvshmem-example.tex
\begin{figure*}
\begin{lstlisting}[caption=Simple one-way bandwidth benchmark with Device-Side GPUSHMEM, label={lst:nvshmem-example}]
//Device-side API 
__global__ void comm_kernel_send(...) {
    nvshmemx_putmem_signal_nbi_block(recv_buf, send_buf, nx, signal_buf + blockIdx.x, 1, NVSHMEM_SIGNAL_ADD, 1);
    grid.sync();
    if (blockIdx.x == 0 && threadIdx.x == 0) 
        nvshmem_signal_wait_until(signal_buf, NVSHMEM_CMP_GE, i + 1);
}

__global__ void comm_kernel_recv(...) { // process 1
    if (threadIdx.x == 0) 
        nvshmem_signal_wait_until(signal_buf + blockIdx.x, NVSHMEM_CMP_EQ, i + 1);
    grid.sync();
    if (blockIdx.x == 0) 
        nvshmemx_putmem_signal_nbi_block(recv_buf, send_buf, 1 * sizeof(real), signal_buf, 1, NVSHMEM_SIGNAL_ADD, 0);
}
//Host-side API 
if (rank == 0) {// process 0
    nvshmemx_collective_launch(comm_kernel_send, dim3(window_size), dim3(1024), kernelArgs, 0, stream);
} else {  // process 1
    nvshmemx_collective_launch(comm_kernel_recv, dim3(window_size), dim3(1), kernelArgs, 0, stream);
}                                                     
\end{lstlisting}
\end{figure*}

%% file: sections/outlook.tex
\section{Discussion, Challenges, and Outlook} \label{Section:Outlook-and-Discussion}
As multi-GPU execution becomes a necessity for many real-world workloads, we dedicate this section to discussions and outlooks, presenting what we believe to be fertile ground for current and future research on GPU-centric communication.


\subsection{\textbf{Moving Away from CPU}}
There are several reasons why we believe moving away from CPU-controlled execution shows promise. First, it addresses the issue of CPU-induced latency barriers that are caused by kernel launch and memory copy overheads. These barriers become more significant in strong scaling scenarios as the number of GPUs increases and computation per GPU decreases. In latency-bound settings, traditional CPU-controlled implementations are not able to overlap communication with computation as the latencies to initiate the operations take longer than the operations themselves effectively serializing communication and computation. On the other hand, CPU-free execution can still achieve adequate levels of overlap even when latencies dominate \cite{Talk-NVSHMEM-Overcoming-Latency-Barriers, Multi-GPU-Communication-Schemes-for-Iterative-Solvers-When-CPUs-Are-Not-in-Charge}. Second, the parallelism offered by within-kernel communication is well-suited for persistent kernels to take advantage of. With the use of efficient communication and synchronization APIs, this execution model can achieve higher bandwidths and lower latencies than CPU-controlled implementations. Additionally, CPU-free execution, by virtue of inlining communication with computation, is well-suited for applications with fine-grained communication \cite{Atos-Vyse-Scalable-Irregular-Parallelism-with-GPUs-Getting-CPUs-Out-of-the-Way}. Third, ROC\_SHMEM allows for the first time to \textit{fully} migrate application execution to the device with zero reliance on CPU helper threads. 
NVSHMEM with its IBGDA transport can also migrate a significant amount of execution to GPU but must still rely on the CPU for functional correctness.

However, there are several challenges facing this model of execution. One major challenge is that persistent kernels can result in reduced occupancy potentially bottlenecking computation. As of today, if global device or multi-GPU barriers are required, persistent kernels must be launched in a \textit{cooperative} manner. This means that only as many threads can be launched as can run concurrently at the same time making hardware oversubscription impossible. As a result, workload decomposition and scheduling, which were previously handled by the hardware scheduler, now need to be manually done by the programmer. This manual approach is unlikely to be as efficient as hardware-based scheduling, and compute-intensive applications are likely to suffer. Furthermore, long-running persistent kernels will consume more registers and may use shared memory as well limiting occupancy even further. Nevertheless, we see a number of solution to this problem. First, there is a large amount of high bandwidth shared memory available across application execution, which can potentially nullify the performance hit caused by reduced occupancy. Second, we predict that as GPU vendors strive more and more for greater GPU autonomy, they will introduce APIs that allow for combining persistent kernels with hardware oversubscription. The manual decomposition can also be handled by an optimized compiler / runtime system. \edit{Alternatively, collective operations can be offloaded to the network components \cite{parallelkittens}, freeing the resources in the GPU to compute.}

Another potential problem with NVSHMEM and ROC\_SHMEM is their ease of integration into existing runtimes. Both libraries center around a \textit{symmetric heap} and all communication buffers must be allocated collectively on the same symmetric heap by all GPUs. This symmetric allocation requires library-specific allocators. Existing runtimes may find it hard to add support for NVSHMEM and ROC\_SHMEM because of the symmetric memory allocation requirement.

\subsection{UCX as a Potential Pathway for GPU-Awareness}
Unified Communication X (UCX) is an open-source communication framework that abstracts over several network APIs, programming models, protocols, and implementations. The idea is to provide a set of high-level primitives while hiding the low-level implementation details behind the UCX runtime. \edit{
UCX is organized into three main components: UCP (UC-Protocol), which provides high-level communication primitives such as messaging and remote memory access; UCT (UC Transport), which offers low-level access to network hardware; and UCS (UC Services), which supplies common utilities for memory management and threading. 

At runtime, a UCP layer dynamically selects the appropriate UCT transport based on the pointer type and system topology. For example, when a CUDA pointer is passed to an inter-node UCP communication, the rc\_mlx5 transport may be chosen, whereas for intra-node communication, the cuda\_ipc transport is typically preferred. As a result, UCP can accept device pointers as data payloads and perform transfers using the most suitable mechanism, making UCX inherently GPU-aware. This GPU-awareness of UCX allows higher-level communication libraries such as OpenMPI and MPICH to handle device pointers directly in their communication primitives. For example in CUDA-aware MPI,  UCX eliminates explicit host memory copies by enabling GPU memory pointers to be passed directly to MPI functions \cite{UCX-an-open-source-framework-for-HPC-network-APIs-and-beyond, openucx-website}. }


Using UCX for realizing \textit{GPU-centric} communication is a recent direction in the literature that has started to take hold. Perhaps the most relevant example is that of \textit{ROCm-awareness} for MPI implementations. Much early work on \textit{GPU-aware} MPI was done for NVIDIA GPUs using native CUDA libraries and then integrated directly into MPI runtimes. Perhaps reticent to replicate the same work to make their runtimes ROCm-aware, most MPI implementations provide ROCm-awareness only through UCX. \edit{OpenMPI and MPICH  additionally provide CUDA support through UCX, besides their native integration. } In non-MPI work, Choi et al. extend the UCX layer in Charm++ to provide GPU-aware communication for several programming models in the Charm++ ecosystem \cite{Charm-GPU-Aware-1-GPU-Aware-Communication-with-UCX-in-Parallel-Programming-Models, Charm-GPU-Aware-2-Accelerating-communication-for-parallel-programming-models-on-GPU-systems}.

We predict that more runtimes will gravitate toward UCX to add support for GPU-centric communication. Using UCX APIs frees the programmers from relying on native vendor-specific APIs and allows adding GPU-aware communication for both ROCm and CUDA. 
\edit{A performance comparison between GPU-awareness through native APIs and that provided through UCX would be helpful. In one work in this direction, Khorassani et al. provide a native ROCm-aware runtime for MVAPICH2, which outperforms OpenMPI with UCX on a cluster of AMD GPUs \cite{Designing-a-ROCm-Aware-MPI-Library-for-AMD-GPUs-Early-Experiences}.} However, the performance difference may be due to differences in the MPI implementations, not UCX.
A current limitation of UCX is that it is always on the host, and it does not use any communication kernel, but relies on the traditional {\em cudamemcpy} family of functions in addition to zero copy RDMA. 





\subsection{CPU-Free Networking} \label{CPU-Free Networking}
As the trend toward GPU-centric communication and greater GPU autonomy continues to accelerate, several works have suggested migrating most or all of the networking stack to the kernel. This is typically done by launching a single long-running persistent kernel and moving both the data and control paths to the GPU.

In the earliest work, GGAS \cite{GGAS-Global-GPU-Adress-Spaces-For-Efficient-Communication-in-Heterogeneous-Clusters} proposes changes to network devices to implement a unified global address space that allows moving the control path entirely to the GPU. This is accomplished by using a persistent kernel that contains the computation, communication, and synchronization all on the device-side. While the work was the first of its kind and showed performance improvements compared to a CUDA-Aware MPI baseline, the experiments were conducted on two GPU nodes with one GPU each, while the proposed hardware changes were emulated on an FPGA. Follow-up work showed that GGAS, by virtue of eliminating CPU involvement in the control path, can achieve further performance improvements and reduce energy usage compared to CPU-controlled baselines \cite{GGAS-Energy-Efficient-Collective-Reduce-and-Allreduce-Operations-on-Distributed-GPUs, GGAS-GPU-Centric-Communication-for-Improved-Efficiency}.

Several more works on CPU-free networking followed. Oden et al. \cite{Infiniband-Verbs-on-GPU-A-Case-Study-of-Controlling-an-Infiniband-Network-Device-from-the-GPU} use GPUDirect RDMA to allow GPUs to directly interface with Infiniband network devices without the involvement of the host CPU. They do this by mapping the entire Infiniband context to the device-side and using the GPU to generate and send work requests to the HCA. However, because of slow single-thread work request generation performance on the GPU, the proposed changes deteriorated performance compared to CPU-controlled baselines. Follow-up work ameliorated these performance limitations and showed much more promising results \cite{Analyzing-Put-Get-APIs-for-Thread-Collaborative-Processors, Analyzing-Communication-Models-for-Distributed-Thread-Collaborative-Processors-in-Terms-of-Energy-and-Time}. Another work combines the proposed GPU-side Infiniband Verbs with CUDA Dynamic Parallelism to optimize the bottleneck of intra-kernel synchronization \cite{Energy-Efficient-Stencil-Computations-on-Distributed-GPUs-Using-Dynamic-Parallelism-and-GPU-Controlled-Communication}. GPUrdma also implements Infiniband on the GPU and proposes a GPU-side library for direct communication from within persistent GPU kernels with zero CPU involvement. The proposed design outperforms a CPU-controlled baseline on a series of microbenchmarks but runs into correctness issues resulting from the use of persistent kernels and GPU-NIC interaction \cite{GPUrdma-GPU-Side-Library-for-High-Performance-Networking-from-GPU-Kenrels}. 

Silberstein et al. implement GPUNet, which provides GPU-side socket abstractions and networking primitives \cite{GPUNet-Networking-Abstractions-for-GPU-Programs}. GPUNet allows invoking the communication on the GPU but does not fully migrate the control path to the device; instead, it relies on CPU helper threads to perform the actual communication. A similar approach is adopted by dCUDA, which provides device-side APIs with MPI semantics but translates them to standard MPI calls performed by CPU helper threads \cite{dCUDA-Hardware-Supported-Overlap-of-Computation-and-Communication}. LeBeane et al. categorized GPU networking methods discussing at length their deficiencies. In response, they propose GPU-TN, a NIC hardware mechanism that allows the CPU to create and register messages with the NIC and the GPU to trigger them from a running persistent kernel \cite{GPU-TN-GPU-Triggered-Networking-for-Intra-Kernel-Communication}. Another work, ComP-Net, uses embedded GPU microprocessors to offload helper threads from the CPU to the GPU \cite{ComP-Net-Command-Processor-Networking-for-Efficient-Intra-Kernel-Communication-on-GPUs}. While both GPU-TN and ComP-Net show promising performance, they require hardware changes to the NIC and the GPU and, thus, rely on simulation to obtain results.


\subsection{Broader GPU Autonomy}
The recent proliferation of \textit{GPU-centric} communication represents the general trend toward broader GPU autonomy. Several works, early and recent, have tried to hand the GPU the reins of domains that have traditionally been the purview of the CPU. In an early work, Stuart et al. propose methods that allow the GPU to issue callbacks to the CPU \cite{GPU-to-CPU-Callbacks}. Silberstein et al. implement GPUfs, which allows the GPU to request files on the host CPU directly from inside a GPU kernel \cite{GPUfs-Integrating-a-File-System-with-GPUs}. Veselý et al. implement support for invoking POSIX system calls from inside GPU kernels through changes to the Linux kernel \cite{Generic-System-Calls-for-GPUs}. NVIDIA's GPUDirect Storage provides a direct data path between GPUs and storage but still relies on the CPU to orchestrate execution \cite{CUDA-Blog-GPUDirect-Storage-A-Direct-Path-Between-Storage-and-GPU-Memory}. 

SmartIO enables low-cost, efficient I/O disaggregation across PCIe-connected hosts by allowing remote machines to borrow and directly access devices such as NVMe, GPUs, and NICs as if they were local.
Building on this idea,  Qureshi et al. at NVIDIA present BaM allowing GPUs to directly access storage without any CPU involvement. Experimental results show that BaM outperforms GPUDirect Storage on several workloads \cite{GPU-Initiated-On-Demand-High-Throughput-Storage-Access-in-the-BaM-System-Architecture}. Turimbetov et al. showed CPU-free device-side task graph execution on multiple GPUs, eliminating kernel launch and scheduling overheads from the host-side \cite{mustard}. Lastly, Baydamirli et al. \cite{Autonomous-Execution-for-Multi-GPU-Systems-Compiler-Support} presented a compiler support for autonomous execution for multi-GPU systems, enabling GPU-initiated communication within Python code using NVSHMEM.  

These and other works show a clear trend toward general GPU autonomy. In line with this, we expect further optimizations to GPU-centric communication. Several recent mechanisms are promising, as pointed out by Punniyamurthy et al. \cite{GPU-initiated-Fine-grained-Overlap-of-Collective-Communication-with-Computation}. First, the recent thread block (TB) cluster abstraction could benefit device-side communication-computation overlap and inter-TB synchronization. Second, AMD's recent cache flush instructions allow flushing the cache before initiating network communication, meaning communication no longer needs to be allocated as uncacheable. Third, recent hardware trends like fatter GPU nodes and tight GPU-NIC integration are also promising \cite{CUDA-Blog-NVIDIA-Grace-Hopper-Superchip-Architecture-In-Depth, AMD-CDNA2-Architecture-Whitepaper}. For example, the most recent iteration of NVSwitch  directly connects 256 Grace Hopper Superchips enabling direct P2P all-to-all communication at an unprecedented scale.

\edit{
While these advances push the system toward broader GPU autonomy, the CPU plays a complementary role to provide system-level responsibilities. Debugging, profiling, monitoring, and software development still fundamentally rely on the CPU’s broader visibility into the system, its mature tooling ecosystem, and its tight integration with operating system services. 
}



\subsection{Collective Algorithms Design}
Multi-GPU systems present unprecedented challenges in the design of collective operations algorithms. First, GPUs within the same node are often arranged in complex and non-traditional topologies, making existing collective algorithms less effective. Second, there is significant heterogeneity in the bandwidth between GPU pairs, both within a node and between nodes. For instance, there can be up to a 4x difference in bandwidth among different GPU pairs on the same node \cite{lumig}, and up to a 10x gap between intra-node and inter-node bandwidth. Third, these topological and connectivity characteristics frequently change between GPU generations. These factors complicate the design of collective algorithms, potentially leading to underutilization of available bandwidth and poor performance of collective operations.

Some approaches use linear programming formulations to find the optimal collective algorithm based on a network specification and the size of the collective \cite{shah2022taccl, 10.1145/3437801.3441620, Blink-Fast-and-Generic-Collectives-for-Distributed-ML}. However, this involves solving an NP-hard problem that scales exponentially. For example, finding a solution for 128 nodes can take up to 11 hours \cite{shah2022taccl}, and a new solution might be needed if the number of nodes or the size of the collective changes. This complexity makes generating collective algorithms for large systems challenging, if not impossible. To simplify the implementation of multi-GPU collective operations, MSCCL \cite{10.1145/3575693.3575724} provides a high-level language for expressing collective operations, which is then compiled to NCCL code.

\subsection{Debugging, Profiling, Benchmarking Support}

Efficient programming tools are essential for productive multi-GPU programming. However, the available tools are severely lacking when it comes to communication native to the GPU. While NVIDIA's flagship system-level profiling tool, NSight Systems, provides a detailed view of host-controlled communication, it falls short in providing information on device-native communication, including Direct Load/Store P2P communication and communication induced by libraries such as NCCL and NVSHMEM.

Palwisha et al. present ComScribe \cite{ComScribe-Identifying-Intra-Node-GPU-Communication}, a tool that allows monitoring NCCL collective and P2P communication. However, it is limited to a single node and relies on the deprecated \textit{nvprof} tool. New efforts in this area have been made by Snoopie \cite{Snoopie}, which focuses on profiling and visualizing GPU-centric communication. Snoopie is capable of attributing communication to source code lines and objects involved in communication, offering different levels of granularity. This allows for both a coarse-grained overview of the system and a detailed view of a specific object or device in terms of data movement. 
Similarly, ucTrace \cite{ucTrace} is a multi-node profiling and visualization framework for the UCX communication layer. It exposes communication behavior across multiple layers of the software and hardware stack.

Another class of tools that can help with debugging GPU communication are race detectors. Given that many of the aforementioned communication libraries use a partitioned global address space model compared with message passing, there is a high likelihood of introducing race hazards. Race detector tools are vital to navigate the complexities of shared data in multi-GPU programming. Despite their importance, none of the existing tools are capable of detecting race hazards in multi-GPU programming. Compute Sanitizer's Racecheck tool \cite{Racecheck} is limited to on-chip shared memory, supporting the detection of race hazards only within a single GPU context. HiRace \cite{HiRace} improves on that by supporting global memory and can detect many types of race hazards that Compute Sanitizer's Racecheck cannot, but it is also limited to a single GPU context.

\edit{Several benchmarking tools have been proposed for measuring the performance of GPU-GPU communication. Standard suites such as the \texttt{*ccl-tests} for NCCL and RCCL are commonly used to assess the performance of both collectives and point-to-point communication~\cite{nccl-tests, rccl-tests}. Similarly, the OSU benchmark suite (OMB) provides extensive support for GPU-aware MPI environments~\cite{10.1007/978-3-642-33518-1_16}. NVSHMEM also provides specific micro-benchmarks to evaluate device-initiated \texttt{put} and \texttt{get} latency. However, these tools typically report performance measurements aggregated across multiple iterations, which masks critical network performance variability and transient noise~\cite{sc2019}. Furthermore, they often lack baselines for explicit peer-to-peer copies. To address these limitations, the Blink-GPU benchmark was proposed~\cite{gpugpuinterconnect, blink-gpu}, enabling the capture of per-iteration timing to expose performance anomalies that aggregated metrics miss.}

We believe that introducing debugging and profiling tools capable of detecting fine-grained device-native transfers both within and across nodes, in addition to race detectors capable of capturing race hazards in a multi-GPU context, is crucial for further advancements in GPU-centric communication.


\subsection{Compression-Accelerated GPU Communication}
\edit{
The high cost of moving large data volumes across the network, particularly for inter-node GPU communication, has motivated the development of techniques for compressing data prior to transmission. While this approach introduces computational overhead for compression and decompression, this cost is frequently amortized by the substantial reduction in communication time. To maximize the data reduction (i.e., achieve a high compression ratio), lossy compression is typically employed. Specifically, error-bounded lossy compression~\cite{10.1145/3733104} allows for high compression ratios while guaranteeing that data distortion remains within a user-defined accuracy threshold. This technique has been integrated into collective communication libraries for both CPU and GPU architectures~\cite{10793125,10.1145/3650200.3656636,10.1145/3721145.3733642,10.1007/978-3-031-07312-0_1, coccl}, demonstrating significant speedups over conventional libraries like MPI and NCCL.

While effective at reducing data volume, these first-generation solutions were fundamentally limited by the Decompression-Operation-Compression (DOC) workflow. This workflow mandates that data be fully decompressed before any computation (such as a reduction in an Allreduce operation) can be performed, only to be immediately recompressed for subsequent steps, incurring significant processing overhead. This limitation has motivated the development of \textit{homomorphic} compression, which enables computation to be performed directly on compressed data. Recent approaches have successfully implemented this paradigm for both CPU~\cite{10793125} and GPU~\cite{10.1145/3721145.3733642} architectures. By allowing GPUs to compute and communicate entirely in the compressed domain, this approach represents the most advanced solution, maximizing throughput by addressing both the data volume bottleneck and the internal DOC processing overhead.
}

\subsection{Level of Maturity and User-Level Perspective}
\edit{It is crucial to clarify the role and maturity of the technologies that underpin modern accelerator communication, such as NVIDIA's GPUDirect, GPUDirect RDMA, AMD's ROCm peer-to-peer capabilities, and GDRCopy. These components are not high-level, user-facing APIs, nor are they \textit{hacks}. Rather, they are foundational, vendor-supported hardware and driver-level functionalities. They represent the official, low-level architecture integrated directly into the accelerator's core runtime environment and the system's kernel drivers. 

The maturity and reliability of these solutions are demonstrated by their universal adoption as the essential building blocks for the entire high-performance, open-source communication ecosystem. High-level libraries, including all major MPI implementations (e.g., Open MPI, MVAPICH, Intel MPI), low-level communication middleware (e.g., UCX), and vendors's own *CCL, are built on top of these APIs. These libraries explicitly use those lower-level functionalities to enable their GPU-aware communication paths.
Therefore, the benefits of these technologies (namely, the elimination of the CPU as a bottleneck and the ability to achieve true, zero-copy, low-latency communication by bypassing host memory) are the well-documented, stable, fundamental mechanisms that enable all modern multi-node, multi-GPU supercomputing.
}

%% file: sections/conclusion.tex
\section{Conclusion} 
\label{Section:Conclusion}

Traditionally, multi-GPU communication was managed by the CPU, but recent advancements in GPU-centric communication have begun shifting this responsibility, allowing GPUs to take more control over the communication tasks. 
This paper provides an in-depth exploration of GPU-centric communication, emphasizing vendor-supplied mechanisms and user-level library support. It seeks to demystify the complexities and variety of options in this area, define key terms, and categorize the various approaches used within individual nodes and across multiple nodes.

The discussion includes an analysis of vendor-provided communication and memory management mechanisms for multi-GPU execution, as well as a review of major communication libraries such as CUDA-aware MPI, NCCL/RCCL/oneCCL, NVSHMEM, ROC\_SHMEM, and Intel SHMEM highlighting their advantages, challenges, and performance considerations. Furthermore, the paper presents important research paradigms such as CPU-free networking, debugging tools for communication, and discusses future directions, and unresolved questions. By thoroughly examining GPU-centric communication techniques across both software and hardware layers, this paper aims to equip researchers, developers, engineers, and library designers with the knowledge needed to fully leverage multi-GPU systems.